\documentclass{article}

\usepackage{arxiv}
\usepackage{amssymb,amsmath,mathtools}
\usepackage{amsmath}
\usepackage{booktabs}
\usepackage{subcaption}
\usepackage{graphicx}
\usepackage{booktabs}
\usepackage{xfrac}
\usepackage{placeins}
\usepackage{array}
\usepackage{longtable,tabularx}
\usepackage{siunitx}
\usepackage{hyperref}
\usepackage{natbib}
\usepackage[utf8]{inputenc}
\usepackage[T1]{fontenc}

\numberwithin{equation}{section}

\newcolumntype{P}[1]{>{\small}{p{#1}}}
\newcolumntype{L}[1]{>{\raggedright\let\newline\\\arraybackslash\hspace{0pt}}m{#1}}
\newcolumntype{C}[1]{>{\centering\let\newline\\\arraybackslash\hspace{0pt}}m{#1}}
\newcolumntype{R}[1]{>{\raggedleft\let\newline\\\arraybackslash\hspace{0pt}}m{#1}}
\newcommand\abs[1]{\left\lvert #1 \right\rvert}
\newcommand\der[2]{\frac{d #1}{d #2}}
\newcommand\pder[2]{\frac{\partial #1}{\partial #2}}
\newcommand\pderv[3]{\left.\frac{\partial #1}{\partial #2}\right\rvert_{#3}}

\newcommand\mbf[1]{\mathbf{#1}}

\title{Towards model predictive control for supercritical CO$_2$ cycles}

\author{
    \href{https://orcid.org/0000-0003-4012-4211}{\includegraphics[scale=0.06]{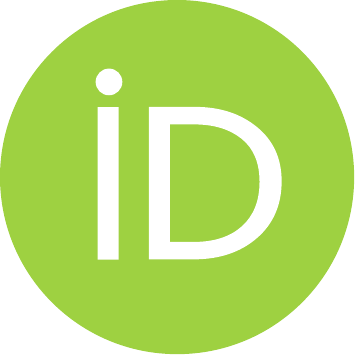}\hspace{1mm}Viv A.~Bone} \\
	School of Mechanical and Mining Engineering\\
	The University of Queensland\\
	St Lucia, QLD, 4072 \\
	\texttt{v.bone@uq.edu.au} \\
	\And
    Michael P.~Kearney \\
	School of Mechanical and Mining Engineering\\
	The University of Queensland\\
	St Lucia, QLD, 4072 \\
	\texttt{m.kearney@uq.edu.au} \\
	\And
    \href{https://orcid.org/0000-0001-6962-8187}{\includegraphics[scale=0.06]{orcid.pdf}\hspace{1mm}Ingo H.\,J.~Jahn} \\
	School of Mechanical and Mining Engineering\\
	The University of Queensland\\
	St Lucia, QLD, 4072 \\
	\texttt{i.jahn@uq.edu.au} \\
}

\renewcommand{\shorttitle}{Towards MPC of sCO$_2$ cycles}

\hypersetup{
pdftitle={Towards model predictive control for supercritical CO2 cycles},
pdfsubject={eess.SY},
pdfauthor={Viv A.~Bone, Michael P.~Kearney, Ingo H.\,J.~Jahn},
pdfkeywords={model predictive control, supercritical co2 cycle},
}

\begin{document}
\maketitle


\begin{abstract}
Key challenges that complicate control of non-condensing non-ideal-gas power
cycles include
(1) their output power dynamics depend on interactions between turbomachinery
and heat transfer processes,
(2) turbomachinery behaviour cannot be captured by simple analytical relations,
and
(3) state constraints must be respected.
This article presents a control methodology for these systems, comprising a
control modelling approach and model predictive control (MPC) strategy.
We demonstrate this methodology on the high-pressure side of a simple
supercritical CO$_2$ cycle power block, composed of a variable-speed compressor,
heat exchanger, and fixed-speed turbine.
We develop a control model by using timescale-separation arguments and locally
linearizing non-ideal-gas turbomachinery performance maps.
We implement MPC by linearizing this control model online at each sampling
instant.
Closed-loop simulations with a full-order gas-dynamics truth model demonstrate
the the effectiveness of the proposed control methodology.
In response to load changes, the controller maintains high turbine inlet
temperatures while achieving net power output ramp rates in excess of 100\% of
nameplate output per minute.
The controller often acts at the intersection of motor torque, compressor surge,
and turbine inlet temperature constraints, and performs well from 35 to 105\% of
nameplate capacity with no parameter scheduling.
The controller achieves good performance and fast update rates when using online
linearization.
The results demonstrate the suitability of MPC for the supercritical CO$_2$
cycle, and the proposed methodology is extensible to more complex cycle variants
such as the recuperated and recompression cycle.
\end{abstract}


\section{Introduction}
\label{sec:introduction}

A critical attribute of thermal power plants is their `flexibility', which is
their ability to operate over a wide range of power outputs and quickly change
from one operating state to another, either by ramping or switching on or off.
Flexibility has four main components:
operating range, part-load efficiency, response times to load changes (i.\,e.\;ramp
rate), and startup and shutdown times and
costs~\citep{feldmuller_2018, irena_conventional_plants_2019}.
Flexible power plants play a crucial roll in maintaining grid reliability under
demand- and supply-side variability~\citep{kassakian_2011}, and they will become
even more important as increasing renewable penetration further increases
supply-side variability and uncertainty~\citep{irena_conventional_plants_2019,
ge_2017, iea_2019}.
Measures of flexibility like minimum load or maximum ramp rate are considered
core performance metrics for modern power plants~\citep{vgb_2017,
agora_2017}.

An impactful way to improve thermal power plant flexibility is to upgrade
plant-level control systems~\citep{ge_2017, vgb_2017, agora_2017}.
Advanced control systems can better coordinate plant subsystems during transients
and facilitate operation closer to component design limits.
These benefits enable operation at higher turn-downs, faster ramping, shorter
and more reproducible start-ups and shut-downs, and decreased
maintenance~\citep{irena_conventional_plants_2019, ge_2017, agora_2017}.
Some examples highlighting the importance of control are as follows:
(1) installation of a new digital control system reduced the minimum load of a
\SI{600}{MW} lignite-fired plant by 27\%~\citep{agora_2017};
(2) for steam plants, control schemes with dynamic wall models have permitted
safe operation closer to thermal stress limits, thus increasing flexibility
without increasing maintenance ~\citep{vgb_2017}; and
(3) commissioning control loops over a wide operating range, rather than only at
nominal load, has significantly improved part-load performance and
ramp rates of coal-fired plants~\citep{vgb_2017}.

Control of thermal power plants, especially in dynamic load-following scenarios,
is challenging for two key reasons:
(1) power plants are multivariable nonlinear systems~\citep{prasad_1998,
prasad_2000, kim_2013, damato_2012, yebi_2017, liu_2017};
(2) power plants are subject to many constraints such as thermal stress
limits~\citep{vgb_2017}, compressor surge~\citep{gravdahl_2012}, and
turbomachinery blade loading limits~\citep{kumar_2012}.
Given these challenges, a suitable control paradigm for this application is
model predictive control (MPC), which performs well for multivariable systems,
systematically handles constraints and plant nonlinearity, and provides a
unified approach to control of complex systems~\citep{maciejowski_2002}.

MPC involves periodically solving forward-looking constrained optimization
problems to compute the control inputs that make a model of the plant best
satisfy the control objectives while respecting all
constraints~\citep{maciejowski_2002}.
Only the first set of inputs is applied, then at the next sampling instant, the
optimization problem is solved again to compute the subsequent set of inputs,
and so on.
The performance of MPC depends strongly on the accuracy and complexity of the
model used by the controller, with more accurate models giving better calibrated
control actions, and simpler models permitting faster updates or longer
prediction horizons. 

Existing literature confirms the value of MPC for established types of thermal
power plants, such as steam Rankine cycle plants~\citep{prasad_1998,
prasad_2000}, open- and closed-cycle gas turbines~\citep{kim_2013, aurora_2005,
damato_2006, damato_2012}, and organic Rankine cycle waste heat recovery
units~\citep{yebi_2017, liu_2017, rathod_2019}.
However, MPC has not yet been investigated for a key
next-generation power cycle: the supercritical carbon dioxide (CO$_2$) Brayton
cycle (the `sCO$_2$ cycle').
This power cycle exploits CO$_2$'s non-ideal-gas behavior
to obtain higher thermal efficiencies than competing power cycles at
readily-achieved turbine inlet temperatures
(600--\SI{700}{\celsius})~\citep{brun_2017_ch1}.
The sCO$_2$ cycle is compact and scalable, being suitable for \SI{10}{MW} rural
plants up to multi-\SI{100}{MW} utility plants~\citep{brun_2017_ch1}.
Additionally, due to the single-phase flow in the heater, the cycle can be well
matched to many heat sources, and CO$_2$'s moderate critical temperature makes
the cycle compatible with non-polluting dry cooling
systems~\citep{turchi_2013}.

However, CO$_2$'s non-ideal-gas behavior gives rise to complex cycle
dynamics~\citep{carstens_2007, moisseytsev_2007}, making the application of
MPC to sCO$_2$ cycles particularly challenging.
Critically, CO$_2$'s strong property variations over the cycle's operating range
complicate modelling of heat exchangers and
turbomachinery~\citep{dostal_2004, carstens_2007, moisseytsev_2007}.
Additionally, most proposed sCO$_2$ cycles use non-condensing
designs~\citep{turchi_2013, rochau_2011}, so their mass flow and pressure
dynamics (which strongly influence power output) arise from complex system
interactions.
Simulation models capture these dynamics implicitly using
quasi-1D compressible flow solvers~\citep{moisseytsev_2007}, but these models
run slower than real time and are thus unsuitable for MPC.
Existing control models for similar systems (such as air Brayton cycles) employ
ideal-gas assumptions~\citep{kim_2013} and thus cannot be used for the
sCO$_2$ cycle.
It is currently unclear how to develop an explicit low-order model of sCO$_2$
cycle dynamics that is suitable for MPC.

This article proposes a control model that can be used to implement MPC for the
high-pressure side of the sCO$_2$ cycle.
The heat exchangers (and other fluid volumes) are modeled using a reduced-order
gas-dynamics model~\citep{bone_2019}, and the turbomachinery are modeled using
performance maps with analytical off-design scaling~\citep{jahn_2017}.
By considering the slow-timescale behavior of the system and linearizing the
turbomachinery maps online, we derive an explicit low-order model for the
slow-timescale evolution of mass flow, pressure, and net output power.
This model can be fitted to individual plants by using modest experimental
datasets to construct turbomachinery performance maps~\citep{glassman_1972} and
tune heat exchanger models~\citep{bone_2018}.
As the model is based primarily on first principles, it is somewhat robust to
large extrapolation or overfitting errors, which may give rise to dangerous
closed-loop behavior.
This model is extensible to commercially-feasible sCO$_2$ cycle configurations
such as the recuperated cycle and recompression cycle.


By linearizing the control model online, we implement MPC for the high-pressure
side of a laboratory-scale open sCO$_2$ cycle, comprising a variable-speed
compressor, printed circuit heat exchanger, fixed-speed turbine, and thermal oil
heat input loop.
The controller's objective is to track net output power setpoints by
manipulating compressor torque and thermal oil flow rate, while also maximizing
turbine inlet temperature and thus cycle thermodynamic efficiency.
The controller must respect the compressor surge constraint and maximum wall
temperature limits.
To assess the performance of the proposed control scheme, we perform closed-loop
simulations with a high-fidelity truth model that is calibrated to experimental
heat exchanger, compressor, and turbine datasets.
This truth model solves the full compressible flow equations and is essentially
a quasi-1D computational fluid dynamics model.
These simulations show the effectiveness of the proposed controller for the test
system, suggesting that the proposed modelling and control strategies are valid.
The results demonstrate MPC's core strengths, namely constraint management, good
dynamic performance for complex multivariable systems, and performance over a
wide nonlinear operating range. 


This article is organized as follows:
Sec.\;\ref{sec:control_problem} defines the control problem considered in
this article;
Sec.\;\ref{sec:simulation_model} presents the simulation model, including the
fluid stream model, heat exchanger model, and the turbomachinery models;
Sec.\;\ref{sec:simulation_model} discusses the derivation of a reduced-order
control model from the simulation model and details how the constraints are
modeled;
Sec.\;\ref{sec:controller_design} discusses the proposed control strategy;
Sec.\;\ref{sec:results_and_discussion} presents the results of closed-loop
simulations; and
Sec.\;\ref{sec:conclusion} concludes the article.



\section{Control problem}
\label{sec:control_problem}

\begin{figure}[h]
\centering
\includegraphics[width=0.55\columnwidth]{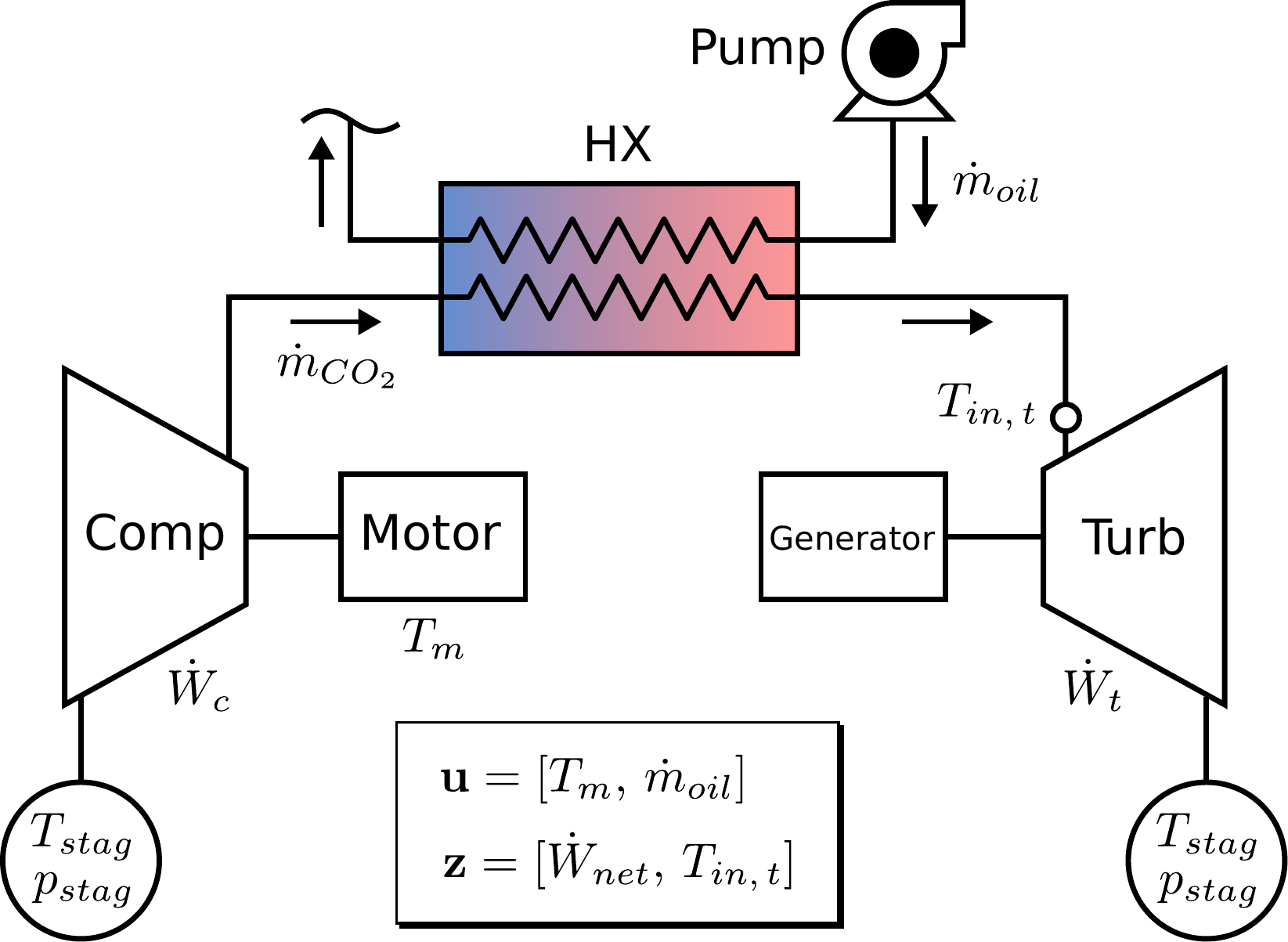}
\caption{Open-cycle system showing inputs $\mbf{u}$ and tracked outputs $\mbf{z}$}
\label{fig:system_schematic}
\end{figure}

This article presents an MPC scheme for the laboratory-scale open sCO$_2$
cycle shown in Fig.\;\ref{fig:system_schematic}.
This system comprises a centrifugal compressor, printed circuit heat exchanger
(PCHE), radial inflow turbine, pump, and connecting pipework.
The process stream working fluid is supercritical CO$_2$, and the heat transfer
fluid is Paratherm\texttrademark\ HE~\citep{paratherm_he}.
The compressor and turbine are on separate shafts.
The compressor is asynchronous and is driven by an electric motor with
controllable torque $T_m$, and the turbine is connected to the grid via a gear
box operating at a synchronous speed of \SI{1200}{RPM}.
A pump drives mass flow in the thermal oil stream.
We assume that a local proportional-integral-derivative (PID) controller
manipulates pump speed to track the thermal oil mass flow rate setpoint
$\dot{m}_{oil,\,ref}$ (the dynamics of this local closed-loop system are given
in Sec.\;\ref{sec:thermal_oil_pump}).
The MPC updates the setpoint of this local PID controller at each sampling
instant.

\begin{figure}[]
\centering
\includegraphics[width=0.60\columnwidth]{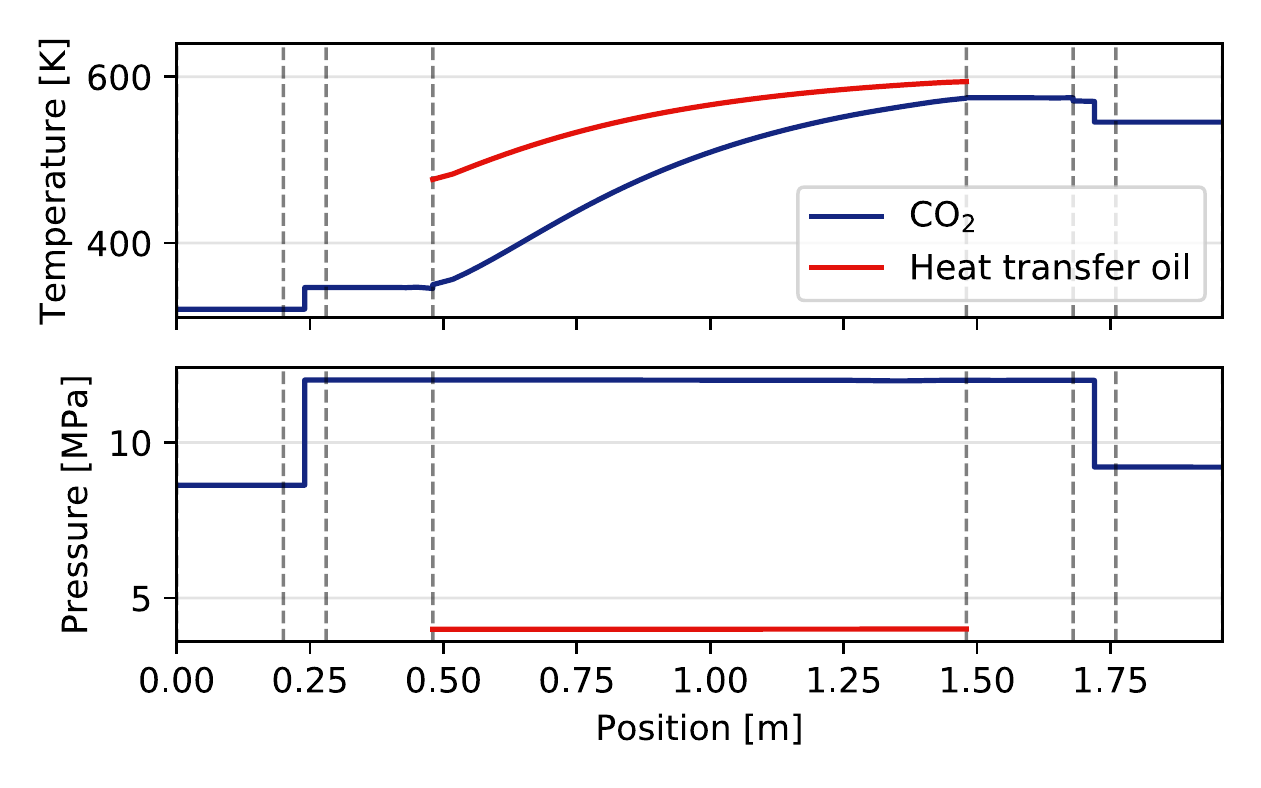}
\vspace{-12pt}
\caption{
Steady-state temperature and pressure profiles --- used as initial conditions
for transient simulations --- showing pipework (I, III, IV,
VI), heat exchanger (IV), compressor (II), and turbine (V).
$\dot{m}_{CO_2} = 10$kg/s, $\dot{m}_{oil} = 10$kg/s.
}
\label{fig:T_p_init}
\end{figure}

The system parameter values are given in Tab.\;\ref{tab:sys_params}.
These parameter values give a nominal power output of \SI{65}{kW}, which is
representative of a laboratory-scale sCO$_2$ cycle, rather than an efficient
commercial facility.
All pipes are assumed to be the same length.
The cell counts $N_{cells}$ refer to the discretization level of the
components in the finite-volume flow solver (see
Sec.\;\ref{sec:simulation_model}).
The compressor rotational inertia represents the compressor rotor plus the
gearbox and motor.
A sample set of steady-state operating conditions, corresponding to a power
output of \SI{55}{kW} and heat input of approximately \SI{3100}{kW}, are shown in
Fig.\;\ref{fig:T_p_init}.

\begin{table}[]
\renewcommand{\arraystretch}{0.9}
\linespread{1.0}\selectfont\centering
{\footnotesize \caption{System parameters\label{tab:sys_params}}
\begin{tabular}{lrllr}
\toprule
\textbf{Heat exchanger}            & \textbf{}                   && \textbf{Pipes}                     & \textbf{}      \\
\midrule                                                          
Length $L$:                        & \SI{1}{m}                   && Length $L$:                        & \SI{0.2}{m}    \\
Channel diameter $d$:              & \SI{1.0}{mm}                && Diameter $d$:                      & \SI{0.08}{m}   \\
Wall thickness $t_w$:              & \SI{1.3}{mm}                && $N_{cells}$ (simulation model):    & 20             \\
Number of channels $N_{chans}$:    & 4000                        &&                                                     \\
$N_{cells}$ (simulation model):    & 100                         &&                                                     \\
\midrule                                                          
\textbf{Compressor}                & \textbf{}                   && \textbf{Turbine}                   & \textbf{}                      \\
Model:                             & See \cite{clementoni_2015}  && Model:                             & Ricardo \& Co A70 \\
Maximum efficiency:                & 0.67                        && Maximum efficiency:                & 0.89                           \\
Design enthalpy rise:              & \SI{25.41}{kJ / kg K}       && Design pressure ratio:             & 1.5                            \\
Design mass flow rate:             & \SI{10}{kg / s}             && Design mass flow parameter:        & 3.2                            \\
Rotational inertia:                & \SI{0.7}{kg m^2}            && Design inlet temperature:          & \SI{600}{K}                    \\
\bottomrule
\end{tabular}
}
\end{table}

The objective of the control scheme presented in this article is regulate the
turbine's power output $\dot{W}_{t}$ to some setpoint $\dot{W}_{t,\,ref}$ while
also driving the turbine inlet temperature to the target value $T_{in,\,t,ref}$.
To maximize the system's thermodynamic cycle efficiency, $T_{in,\,t,ref}$ is set
as the maximum feasible turbine inlet temperature that can be achieved for
$\dot{W}_{t,\,ref}$.
The control inputs are $\dot{m}_{oil,ref}$ and $T_m$.

\begin{figure}[]
\centering
\includegraphics[width=0.55\columnwidth]{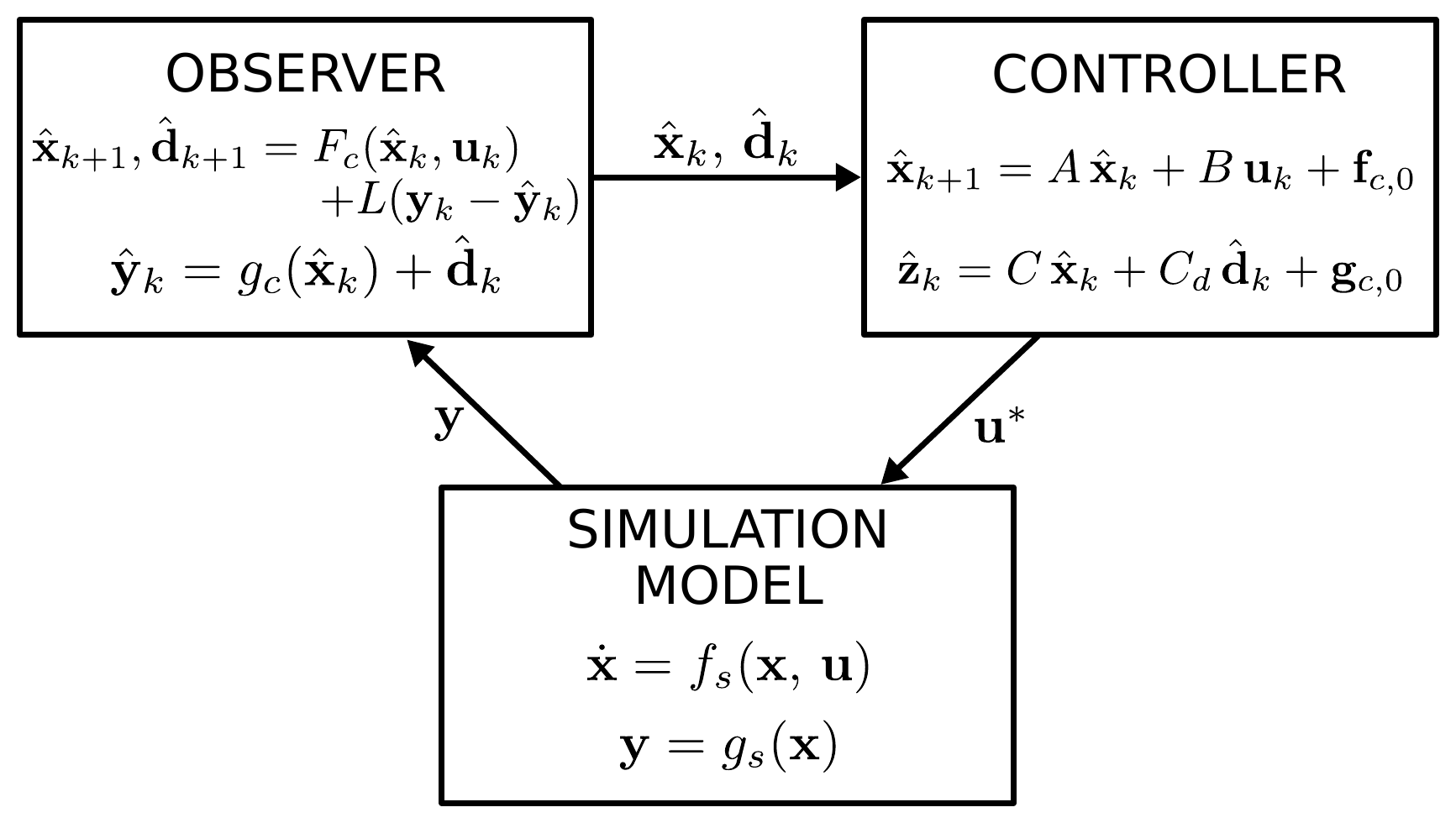}
\caption{Simulation and control approach (detailed in Sec.~\ref{sec:controller_design})}
\label{fig:simulation_setup}
\vspace{-18pt}
\end{figure}

We analyze the performance of the proposed MPC scheme using the closed-loop
simulation setup shown in Fig.\;\ref{fig:simulation_setup}.
This setup uses a high-fidelity simulation model (see
Sec.\;\ref{sec:simulation_model}) with states $\mbf{x}$ and measured outputs
$\mbf{y}$ as a substitute for the real plant.
At each sampling instant, the control update procedure is as follows:
First, the observer, based on the reduced-order control model
(see Sec.\;\ref{sec:control_model}), generates state estimates $\hat{\mbf{x}}$
and disturbance estimates $\hat{\mbf{d}}$ from measurements $\mbf{y}$.
The controller linearizes the control model about the current operating point
($\hat{\mbf{x}}_{0}$, $\mbf{u}_0$), then uses this linear model to compute the
optimal inputs $\mbf{U}^{\ast}$ that drive the tracked outputs $\mbf{z}$ to
their target values.
Finally, the first set of optimal inputs $\mbf{u}^{\ast}$ are applied to the
simulation model.


\section{Simulation model}
\label{sec:simulation_model}

sCO$_2$ cycles are an early-stage technology and expensive to construct,
so development of new control strategies on real plants is impractical.
Accordingly, we develop and test our control strategy using a high-fidelity
simulation model that incorporates experimental heat exchanger, compressor, and
turbomachinery datasets.
Our simulation model is developed using a similar approach as the Argonne
National Laboratory Plant Dynamics Code~\citep{moisseytsev_2007}, which is
a thoroughly validated~\citep{moisseytsev_2011, moisseytsev_2012,
moisseytsev_2016} dynamic model for sCO$_2$ cycles (and is not publicly
available).
This section presents the simulation model, which is developed by combining
submodels of five component types --- pipes, heat exchangers, turbomachinery,
pumps, and boundaries --- in the configuration shown in
Fig.\;\ref{fig:system_schematic}.



\subsection{Quasi-1D streams}
\label{sec:quasi_1d_streams}

Submodels for pipes, heat exchangers, and turbomachinery are developed using the
quasi-1D stream as a fundamental building block.
Being \emph{quasi}-1D, these streams account for the effects of cross-sectional
geometry on flow velocity, heat transfer, and pressure drop, and so can be used
model components with 3D axially-dominated flow~\citep{carstens_2007,
moisseytsev_2007}.

We neglect axial conduction within quasi-1D streams as it is negligible for the
component types of interest.
In printed heat heat exchangers, transverse thermal gradients are much greater
than axial ones, so axial conduction in the fluid can safely be neglected (as
shown experimentally in~\citep{bone_2018}).
And in turbomachinery, the high axial velocities of the fluid mean that axial
conduction is negligible.
Under this assumption, the governing equations for quasi-1D flow with heat
transfer and frictional pressure drop are
\begin{subequations}
\label{sys:q1d_flow_eqs}
    \begin{alignat}{2}
    \label{eq:1DFlowContinuity}
    A\, \frac{\partial \rho}{\partial t}     &= - \frac{\partial (\rho v A)}{\partial x}
    \\
    \label{eq:1DFlowMomentum}
    A\, \frac{\partial (\rho v)}{\partial t} &= - \frac{\partial \left(\rho v \abs{v} A + p A\right)}{\partial x} - p \pder{A}{x} - \text{fr}\, \frac{\rho v \abs{v} A}{2 D_H}
    \\
    \label{eq:1DFlowEnergy}
    A\, \frac{\partial (\rho E)}{\partial t} &= - \frac{\partial (\rho H v A)}{\partial x} - \text{fr}\, \frac{\rho v^3 A}{2 D_H} - q''.
    \end{alignat}
\end{subequations}
%
where $t$ is time, $x$ is the spatial coordinate, $v$ is velocity, $p$ is
pressure, $\rho$ is density, $e$ is specific internal energy, specific enthalpy
is $h = e + p/\rho$, specific total energy is $E = e + \frac{1}{2}v^2$, specific
total enthalpy is $H = h + \frac{1}{2}v^2$, $D_H$ is the hydraulic diameter, fr
is the Darcy friction factor, $A$ is the total flow area, and $q''$ is the heat
flux through the wall.
These equations are closed using an equation of state for the relevant fluid
(see Sec.\;\ref{sec:fluid_property_calculations}).
To solve, we integrate Eqs.\;\ref{sys:q1d_flow_eqs}
over $N_{cells}$ finite volumes spanning the computational domain, yielding a
system of temporal ODEs for the evolution of density, velocity, and total energy
in each volume.

For incompressible working fluids, density and pressure are independent, so the
discretized continuity equation (Eq.\;\ref{eq:1DFlowContinuity}) becomes a
constraint on the velocity field, rather than a transport equation for
density~\citep{versteeg_2007_ch6}.
In this case, at each timestep, we use the unsteady PISO
algorithm~\citep{issa_1986} with implicit Euler time integration to compute the
pressure and velocity fields, and the unsteady energy equation with central
differencing to compute the temperature field (consult \citep{issa_1986} or
\citep{versteeg_2007_ch6} for details).
The remaining fluid properties are computed from temperature and pressure using
the equation of state.
We assume that fluid properties are fixed at the inflow boundary and that the
flow is fully-developed at the outflow boundary.

For compressible working fluids, we directly solve the governing temporal ODEs
using the AUSMDV flux splitting scheme~\citep{wada_1997} with 4$^{th}$-order
Runge-Kutta Cash-Karp time integration~\citep{cash_1990} (refer to the
sources for details).
To accurately model pressure and mass flow dynamics in fluid streams that
are driven by turbomachinery, compressible boundaries must be modeled properly.
To replicate the physics of a real system, we model these boundaries by assuming
that the fluid flows from some infinitely-large inlet reservoir, through the
computational domain, then into an infinitely-large outlet reservoir (see
Sec.\;\ref{sec:compressible_flow_boundary_conditions}).


\subsubsection{Compressible-flow boundary conditions}
\label{sec:compressible_flow_boundary_conditions}

\begin{figure}[h]
\centering
\includegraphics[width=0.50\columnwidth]{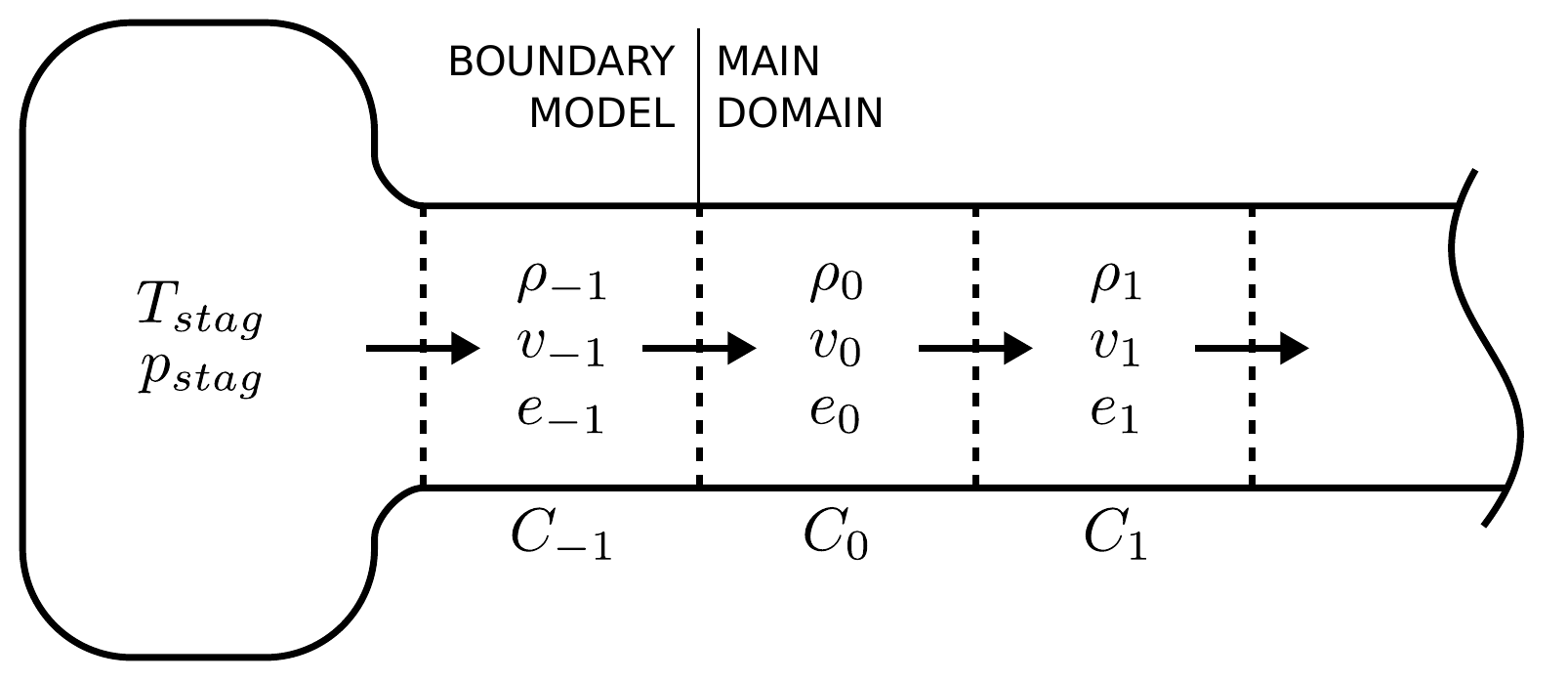}
\caption{Inflow boundary model}
\label{fig:stagnation_boundary}
\end{figure}

For compressible fluids, inflow boundaries are modeled by assuming that the
fluid isentropically accelerates from a reservoir at some fixed stagnation
conditions $(p_{stag,\,in},\,T_{stag,\,in})$ into the computational domain
(similar to the subsonic boundary model in~\citep{jacobs_2017}).
Fig.\;\ref{fig:stagnation_boundary} shows a schematic of the inflow boundary
model.

At beginning of each timestep, the boundary model sets the property values in
$C_{-1}$, allowing $C_0$ may be treated normally by the compressible flow
solver.
$C_{-1}$ is updated as follows.
We assume zero velocity gradient over the inflow boundary, and so set
\begin{equation}
\label{eq:inflow_bc_start}
v_{-1} = v_0
\end{equation}
where $v_{-1}$ and $v_{0}$ are the velocities in cells $C_{-1}$ and $C_0$
(and $v_0$ was computed during the previous timestep). 
Assuming that the fluid accelerates isentropically from the reservoir into
$C_{-1}$, the enthalpy $h_{-1}$ is
%
\begin{equation}
h_{-1} = h_{stag,\,in} - \frac{v_{-1}^2}{2}
\end{equation}
where $h_{stag,\,in} = \text{EOS}(p_{stag,\,in},\,T_{stag,\,in})$, and
the notation $w = \text{EOS}(\alpha,\, \beta)$ means to compute the property $w$
from properties $\alpha$ and $\beta$ using an appropriate equation of state (see
Sec.\;\ref{sec:fluid_property_calculations}).
From the stagnation entropy, $s_{stag,\,in} =
\text{EOS}(p_{stag,\,in},\,T_{stag,\,in})$,
all other properties in $C_{-1}$ are computed using the equation of state.

For outflow boundaries, we assume that the fluid decelerates from the final cell
into the reservoir via a completely irreversible process where all kinetic
energy is dissipated.
This assumption means that the pressure in the final cell is fixed to the
reservoir pressure $p_{stag,\,out}$ and the temperature is set by the outgoing
fluid temperature.
As for the inflow, we compute the outflow velocity using a zero-gradient
approximation.


\subsection{Pipework}
\label{sec:pipework}

Pipes are modeled as quasi-1D streams (Eqs.\;\ref{sys:q1d_flow_eqs})
without heat transfer.
The well-established Darcy and Colebrook-White~\citep{white_2011} formulae are
used to compute the friction factor for laminar and turbulent flow respectively.


\subsection{Heat exchangers}
\label{sec:heat_exchangers}

This article focuses on PCHEs as they are most suitable for the heat addition
and recuperation processes in sCO$_2$ cycles~\citep{brun_2017_ch8}.
PCHEs consist of several layers of metal plates into which zigzag-shaped
microchannels have been chemically etched.
Alternating hot-stream and cold-stream plates are arranged on top of one
another in a counter-flow configuration, then diffusion bonded together to
create a solid block~\citep{brun_2017_ch8}.

\begin{figure}[h]
\centering
\includegraphics[width=0.60\columnwidth]{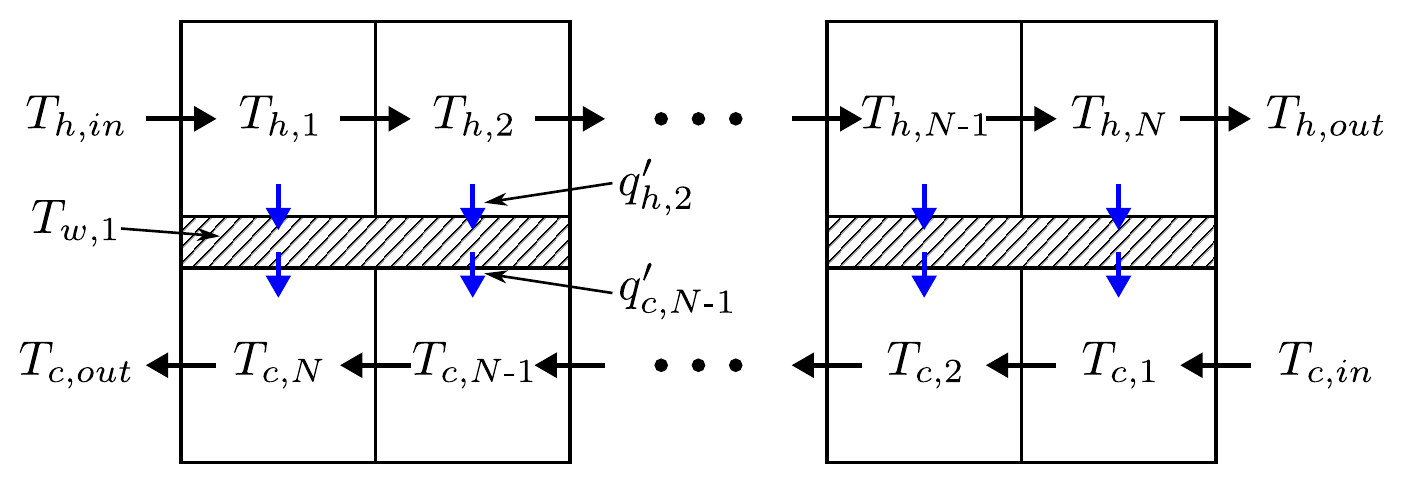}
\caption{PCHE modelling approach}
\label{fig:pche_modelling_approach}
\end{figure}

We develop a plant-level simulation model of a PCHE by following the approach
of~\citet{carstens_2007} and \citet{moisseytsev_2007}.
We assume that
1) mass flow is distributed evenly between all channels,
2) fluid properties in all channels are the same at a given axial distance into
the heat exchanger, and
3) all channels have the same geometry.
Under these assumptions, PCHEs may be modeled as two representative 1D fluid
channels separated by a conductive wall, as illustrated in
Fig.\;\ref{fig:pche_modelling_approach}.
Each representative channel is modeled using the quasi-1D flow equations
(Eqs.\;\ref{sys:q1d_flow_eqs}), where the flow area and heat transfer area
$A_w$ are scaled by the number of channels $N_{chans}$.
We use the Colebrook-White~\citep{white_2011} formula to compute the friction
factor for turbulent flow.

Because (1) axial conduction in PCHE walls is minimal and (2) the thermal
resistance for cross-wall conduction is small compared to that for forced
convection~\citep{bone_2018}, we assume that heat flows through the heat
exchanger walls strictly in the transverse direction (see
Fig.\;\ref{fig:pche_modelling_approach}).
Under these assumptions, wall temperature dynamics are given by
\begin{equation}
\label{eq:1d_wall_energy}
A_{w}\, \rho_{w}\, C_{p,\, w}\, \der{T_{w}}{t} = q''_{h} + q''_{c},
\end{equation}
where $\rho_{w}$, $C_{p,\, w}$, and $T_w$ are the density, heat capacity, and
mean wall temperature in the transverse direction respectively.
$q''_{h}$ is the heat flux to the hot stream, given by
\begin{equation}
q''_{h} = N_{chans} U_h P_h (T_{h} - T_{w})
\end{equation}
where for the hot channel, $T_h$ is the fluid temperature, $P_h$ is the wetted
perimeter and $U_h$ is the local heat transfer coefficient (the cold stream heat
flux $q''_{c}$ is treated analogously).
The heat transfer coefficient is computed using the Nusselt number Nu, the
channel's characteristic length $L_C$, and the fluid's thermal conductivity $k$
as, for the hot stream,
\begin{equation}
\label{eq:heat_transfer_coefficient}
U_h = \text{Nu}_h\, k_h / L_{C,\,h}.
\end{equation}
To accurately model heat exchanger behavior, it is crucial to select valid heat
transfer correlations that properly account for the effects of channel geometry
and multi-dimensional fluid flow~\citep{bone_2018}.


\subsubsection{Heat transfer correlations}
\label{sec:heat_transfer_correlations}

Due to its property variations, supercritical CO$_2$ exhibits complex heat
transfer behavior, especially near its critical point.
Accurately modelling supercritical CO$_2$ heat transfer over a wide operating
range likely requires several switched heat transfer correlations.
However, as this work focuses on control design, we select a heat transfer
correlation that gives representative heat transfer behavior of a supercritical
CO$_2$ PCHE near its design point.
(Our approach is trivially extensible to cases with switched heat transfer
correlations.)
Moreover, by fitting the heat exchanger model to small experimental datasets
collected from real heat exchangers, heat transfer behavior can be modeled
very accurately with only approximate knowledge of internal geometry or heat
transfer correlations~\citep{bone_2018}.

Flow in the PCHEs in the sCO$_2$ cycle is typically turbulent, so we use the
Ngo-Ishizuka correlation~\citep{ngo_2007} to model supercritical CO$_2$ heat
transfer.
This correlation was developed for turbulent flow in channels of hydraulic
diameter \SI{1.09}{mm} and zigzag angle \SI{52}{\degree} using experimental
data, and gives Nusselt number as
\begin{equation}
\label{eq:ngo}
\text{Nu} = 0.1696\, \text{Re}^{0.629}\, \text{Pr}^{0.317},
\end{equation}
where Re is the Reynolds number and Pr is the Prandtl number.

The thermal oil flow is laminar with Re typically less than 200.
Accordingly, its heat transfer is unlikely to be affected by the zigzag channel
geometry, so we model its heat transfer using the analytical
correlation for fully-developed laminar flow in a straight semi-circular duct
with constant heat flux and transversely invariant
properties~\citep{faghri_2010}:
\begin{equation}
\label{eq:oil_nu}
\text{Nu} = 4.089
\end{equation}


\subsection{Turbomachinery}
\label{sec:turbomachinery}

Turbomachinery exhibit complex thermo-fluid behavior that cannot be modeled in
detail in a plant-level simulation model.
Here, we develop a suitable turbomachinery model following a similar approach
to \citet{moisseytsev_2007}.
This approach assumes that the thermodynamic response of turbomachines is
instantaneous, so their outlet thermodynamic state ($p_{out}$, $T_{out}$) is a
function $f_{tb}$ of only their inlet fluid state ($p_{in}$, $T_{in}$), mass flow rate
$\dot{m}_{tb}$, and shaft speed $N_s$:
\begin{equation}
\label{eq:performance_map}
p_{out},\, T_{out} = f_{tb}(p_{in},\, T_{in},\, \dot{m}_{tb},\, N_s).
\end{equation}
%
Under this assumption, turbomachines may be modeled using either simplified
meanline models or performance maps (which may be generated from experiments or
CFD~\citep{jahn_2017}).
This work uses 2D performance maps with analytical off-design scaling
(detailed in Appendix \ref{sec:performance_maps}).
However, the control strategy proposed in this article works with any 
turbomachinery model that fits the form of Eq.~\ref{eq:performance_map}.
When modelling an operational plant, highly-accurate 4D performance maps would
likely be used since detailed operating data would be available.


\subsubsection{Integration of turbomachinery models into the flow solver}
\label{sec:integration_of_turbomachinery_models_into_the_flow_solver}

%

\begin{figure}[hb]
\centering
\includegraphics[width=0.55\columnwidth]{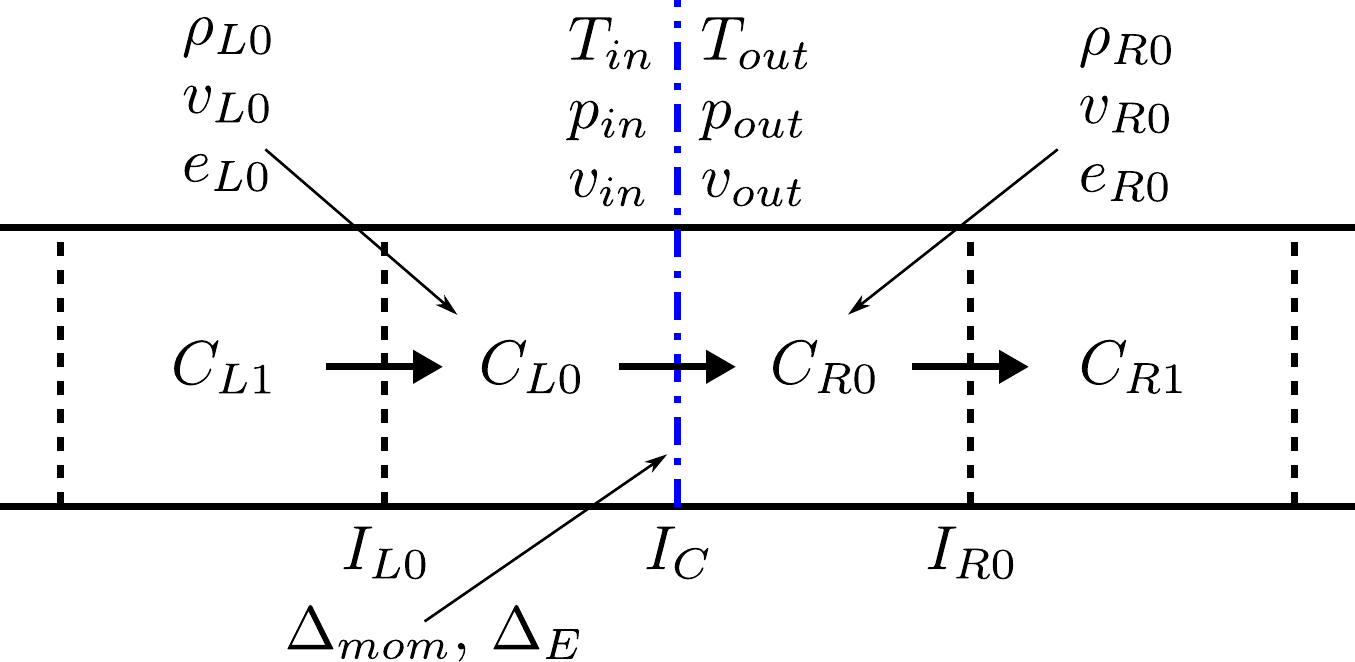}
\caption{Turbomachinery modelling concept}
\label{fig:turbomachinery_modelling_concept}
\end{figure}

Using performance maps, turbomachines can be modeled as two adjacent
compressible flow cells ($C_{L0}$ and $C_{R0}$) with momentum and energy
discontinuities $\Delta_{mom}$ and $\Delta_E$ over the interface $I_C$ that
separates them (see Fig.\;\ref{fig:turbomachinery_modelling_concept}).
With these discontinuities, the flux vectors on the left and right of $I_C$
($F_{L}$ and $F_{R}$ respectively) are
%
\begin{equation}
\begin{bmatrix}
F_{m,\,R}
\\
F_{mom,\,R}
\\
F_{E,\,R}
\end{bmatrix}
=
\begin{bmatrix}
F_{m,\,L}
\\
F_{mom,\,L} + \Delta_{mom}
\\
F_{E,\,L} + \Delta_E
\end{bmatrix}
\end{equation}
where the subscripts $_m$, $_{mom}$, and $_E$ refer to mass, momentum, and energy.
During gas dynamic updates, cell $C_{L0}$ is integrated using $F_L$ and the flux
vector at $I_{L0}$, similarly $C_{R0}$ is integrated using $F_R$ the flux vector at
$I_{R0}$.
The discontinuities $\Delta_{mom}$ and $\Delta_E$ are not directly computed;
they emerge from the process used to compute $F_L$ and $F_R$ from the
turbomachinery maps.

Because (1) mass flow rate through a turbomachine is primarily governed by pressure
ratio and (2) off-design scaling is performed based on the inlet fluid
state, we reformulate the performance maps as a function of pressure
ratio~\citep{moisseytsev_2007}:
\begin{equation}
\label{eq:performance_map_reformulated}
T_{out},\, \dot{m}_{tb} = f_{tb}'(p_{in},\, p_{out},\, T_{in},\, N_s).
\end{equation}
%
Using these reformulated performance maps, $F_L$ and $F_R$ are calculated as
follows.
First, the shaft speed is updated from rotordynamics (see
Appendix Sec.\;\ref{sec:turbomachinery_rotor_dynamics}).
Next, from the current fluid states in $C_{L0}$ and $C_{R0}$ we set
%
%
\begin{equation}
T_{in}  = T_{L0}, \quad
p_{in}  = p_{L0}, \quad
p_{out} = p_{R0}.
\end{equation}
Then, we compute $T_{out}$ and $\dot{m}_{tb}$ from the performance maps
(see Appendix~\ref{sec:performance_maps}),
then compute the inlet and outlet
velocities $v_{in}$ and $v_{out}$ and shaft work $\dot{W}$ as
\begin{align}
v_{in}      &= \dot{m}_{tb} \; / \left( A_{tb,\,in}\, \rho_{in} \right),
\\
v_{out}     &= \dot{m}_{tb} \; / \left( A_{tb,\,out}\, \rho_{out} \right),
\\
\label{eq:w_s}
\dot{W} &= \dot{m}_{tb}\, \left(h_{out} - h_{in} + \sfrac{1}{2}\, \left(v_{out}^2 - v_{in}^2\right) \right),
\end{align}
where $A_{tb,\,in}$ and $A_{tb,\,out}$ are the inlet and outlet flow areas of
the turbomachine.
This solution approach implicitly captures efficiency from the turbomachinery
maps.
Finally, the flux vectors are formed as
\begin{align}
F_L &=
\begin{bmatrix}
\dot{m}_{tb} / A_{tb,\,in}
\\
\left(\dot{m}_{tb}\, v_{in} / A_{tb,\,in}\right) + p_{in}
\\
\left(\dot{m}_{tb}\, E_{in} / A_{tb,\,in}\right)
\end{bmatrix}
\end{align}
and
\begin{align}
F_R &=
\begin{bmatrix}
\dot{m}_{tb} / A_{tb,\,out}
\\
\left(\dot{m}_{tb}\, v_{out} / A_{tb,\,out}\right) + p_{out}
\\
\left(\dot{m}_{tb}\, E_{in} / A_{tb,\,in}\right) - \dot{W}_s
\end{bmatrix}.
\end{align}
When evaluating interfaces fluxes for $I_{L0}$ and $I_{R0}$, ghost cells
populated with first-order extrapolation are used to prevent the compressible
flux calculator from `looking over' the momentum and energy discontinuity at
$I_C$.




\subsubsection{Turbomachinery rotor dynamics}
\label{sec:turbomachinery_rotor_dynamics}

Turbomachines may be either synchronous (grid-connected at some constant speed
$N_{s0}$) or asynchronous (variable-speed).
The rotational dynamics of asynchronous turbomachines depend on the
load and external torques ($T_{load}$ and $T_{external}$).
For compressors, $T_{external}$ is the applied motor torque, and for
turbines, $T_{external}$ is the supplied generator torque.
The load torque is
\begin{equation}
T_{load} = \dot{W}_s / N_s,
\end{equation}
where $\dot{W}_s$ is computed using performance maps
(see Appendix~\ref{sec:performance_maps})
and Eq.\;\ref{eq:w_s}.
(see Appendix~\ref{sec:performance_maps}),
Using these torques, turbomachinery rotational dynamics are given by
\begin{equation}
\label{eq:rotordynamics}
J \frac{dN_s}{dt} = T_{external} - T_{load}.
\end{equation}
where $J$ is the moment of inertia of the turbomachine's rotor plus connected
rotating masses (such as gearbox components).






\subsection{Thermal oil pump}
\label{sec:thermal_oil_pump}

The mass flow rate in the thermal oil stream $\dot{m}_{oil}$ is set by a pump.
We assume that the fluid exits the pump at $T_{oil,\,in} = $ \SI{300}{\celsius}
and $p_{oil,\,in} = $ \SI{4}{MPa}, and we assume that a PID controller modulates
pump speed such that $\dot{m}_{oil}$ is driven towards the setpoint
$\dot{m}_{oil,\,ref}$.
We assume that the local closed-loop dynamics of the pump and controller
subsystem are
%
\begin{equation}
\label{eq:pump_dynamics}
J_{p}\, \frac{d^2 \dot{m}_{oil}}{d t^2} = k_{p}\, (\dot{m}_{oil,\,ref} - \dot{m}_{oil}) - c_{p}\, \frac{d \dot{m}}{d t},
\end{equation}
where $J_p$, $c_p$, and $k_p$ are the effective inertia, damping coefficient, and
stiffness respectively.
We assume that the PID gains are set such that the pump and controller subsystem
has a natural frequency of \SI{2}{Hz} and slightly overdamped response, with
damping ratio 1.3.


\subsection{Fluid property calculations}
\label{sec:fluid_property_calculations}

We form fluid property lookup tables using the open-source fluid property
database \verb'CoolProp'~\citep{CoolProp}.
\verb'CoolProp' computes CO$_2$'s thermodynamic state by iteratively solving
the Span and Wagner equation of state~\citep{span_1996}, and computes its thermal
conductivity and viscosity using the Scalabrin et.\;al.\;~\citep{scalabrin_2006}
and Fenghour et.\;al.\;~\citep{fenghour_1998} correlations respectively.
\verb'CoolProp' models Paratherm\texttrademark HE using an incompressible
equation of state based on manufacturer data~\citep{paratherm_he}.


For computational efficiency, we compute fluid properties from the lookup tables
using bicubic interpolation.
Discontinuous approximation methods (such as tabular Taylor series
extrapolation) make the compressor performance map model used in this article
ill-conditioned and thus cannot be used.



\section{Control model}
\label{sec:control_model}

The simulation model presented in Sec.\;\ref{sec:simulation_model} is
ill-suited for control applications due to its high state dimension and
stringent explicit timestep restrictions, imposed by the
Courant-Friedrichs-Lewy condition~\citep{courant_1967}.
This section discusses the formulation of a simpler control model of the open
sCO$_2$ cycle (detailed in Sec.\;\ref{sec:control_problem}).
This control model (denoted by $f_c$ and $g_c$ in
Fig.\;\ref{fig:simulation_setup}), is linearized online to implement MPC and is
also used to estimate unmeasured states.
The control model is primarily composed of two submodels: one for the thermal
and bulk-flow dynamics in non-ideal fluid streams (used for both the CO$_2$ and
thermal oil streams), and one for the slow-timescale pressure and mass flow
dynamics in the CO$_2$ stream.

Previous work~\citep{bone_2019} shows that by assuming the momentum dynamics in
compressible fluid streams are infinitely-fast (the `quasi-steady-momentum'
assumption), these streams can be modeled using quasi-incompressible flow
equations, which capture only the slow-timescale energy and bulk-flow dynamics.
This work shows that the quasi-steady-momentum assumption can also be used to
derive a control model for the pressure and mass flow dynamics in compressible
flow systems that are driven by turbomachinery behavior and heat transfer
dynamics.


\subsection{Turbomachinery rotor dynamics}
\label{sec:turbomachinery_rotor_dynamics_linearized}

In the control model, the rotordynamics of asynchronous turbomachines are
modeled by linearizing Eq.\;\ref{eq:rotordynamics} about the current
operating point.
The load torque is computed using the performance map model, and so can be
considered as a function $T_{load}(p_{in},\, p_{out},\, T_{in},\, N_s)$.
For the compressor, with $T_{external}$ as the motor torque $T_m$, the
linearization of Eq.\;\ref{eq:rotordynamics} is
\begin{equation}
\begin{split}
\label{eq:rotordynamics_linearized}
J_{c} \der{N_{c}}{t} \ = \ & T_{m,\,0} + \overline{T}_m - T_{load,\,0} - \pder{T_{load}}{N_{c}}\, \overline{N}_{c}
\\
                           & - \pder{T_{load}}{\dot{m}_{c}}\, \overline{\dot{m}}_{c} - \pder{T_{load}}{T_{c,\,in}}\, \overline{T}_{c,\,in} - \pder{T_{load}}{p_{c,\,in}}\, \overline{p}_{c,\,in}.
\end{split}
\end{equation}
where the `overbar' variables represent deviation from the linearization point
and `subscript 0' variables denote values at the linearization point
(see Sec.\;\ref{sec:linearization_and_discrete-time_conversion}).
As discussed in in Sec.\;\ref{sec:mass_flow_rate_and_pressure_modelling}, in the
control model, the compressor inlet temperature and pressure are considered
fixed, so Eq.\;\ref{eq:rotordynamics_linearized} can be simplified to
\begin{equation}
\label{eq:rotordynamics_linearized_simplified}
J_{c} \der{N_{c}}{t} = T_{m,\,0} + \overline{T}_m
- T_{load,\,0} - \pder{T_{load}}{N_{c}}\, \overline{N}_{c} - \pder{T_{load}}{\dot{m}_{c}}\, \overline{\dot{m}}_{c}.
\end{equation}


\subsection{Heat exchanger walls}
\label{sec:heat_exchanger_walls}

The simulation model for the heat exchanger walls is appropriate for
control without further simplification.
Integrating Eq.\;\ref{eq:1d_wall_energy} over $N_{cells}$ volumes yields the following
temporal ODE for temperature in wall cell $i$:
%
\begin{equation}
\begin{split}
\label{eq:wall_temp_update}
A_{w}\, \rho_{w}\, C_{p,\, w}\, \der{T_{w,\, i}}{t} \ = \ & N_{chans}\, (U_h\, P_{C,\, h} (T_{h,\, i} - T_{w,\, i})
\\
                                                          & + U_c\, P_{C,\, c} (T_{c,\, i} - T_{w,\, i}))
\end{split}
\end{equation}
where $T_{h,\, i}$, $T_{c,\, i}$ are the neighboring hot-stream and cold-stream
fluid cells, and the heat transfer coefficients $U_c$ and $U_h$ are given by
Eq.\;\ref{eq:heat_transfer_coefficient}.
Eq.\;\ref{eq:wall_temp_update} is linearized online to implement MPC (see
Sec.\;\ref{sec:linearization_and_discrete-time_conversion}).
Best closed-loop performance was obtained by retaining only the `primary'
partial derivatives in the linearization, giving
\begin{equation}
\label{eq:wall_temp_update_linear}
\begin{split}
\der{T_{w,\, i}}{t} \; \approx \; & \frac{N_{chans}}{A_{w}\, \rho_{w}\, C_{p,\, w}} \left( \der{T_{w,\, i,\, 0}}{t} - (U_h\, P_{C,\, h} + U_c\, P_{C,\, c}) \overline{T}_{w,\,i} \right.
\\
                                & \left. + U_h\, P_{C,\, h}\, \overline{T}_{h,\, i} + U_c\, P_{C,\, c}\, \overline{T}_{c,\, i} \vphantom{\der{T_{w,\, i,\, 0}}{t}} \right),
\end{split}
\end{equation}
where $\der{T_{w,\, i,\, 0}}{t}$ is the derivative of wall temperature at the
linearization point.


\subsection{Fluid streams}
\label{sec:fluid_streams}


For control, we model only the slow-timescale dynamics (the thermal
and bulk-flow dynamics) of quasi-1D streams~\citep{bone_2019} by applying
the quasi-steady momentum assumption,
\begin{equation}
\label{eq:quasi_steady_momentum}
\pder{(\rho v)}{t} \approx 0,
\end{equation}
to the quasi-1D flow equations (Eqs.\;\ref{sys:q1d_flow_eqs}).
This assumption effectively states that the momentum dynamics (i.\,e.\; pressure
waves) stabilize infinitely quickly compared to the other dynamics.
For the operating conditions considered in this article, this assumption is
justified as Mach numbers in the CO$_2$ stream are always less than 0.01,
so the pressure wave characteristics are much faster than the bulk-flow
characteristic~\citep{bone_2019}.
Similarly low Mach number ranges have been observed in experimental
tests of supercritical CO$_2$ heat exchangers~\citep{bone_2018}.

Additionally, we make two more assumptions:
(1) that the pressure changes throughout the fluid streams are
negligible except over the turbomachinery (see
Sec.\;\ref{sec:mass_flow_rate_and_pressure_modelling}), and
(2) that $H \approx h$ and $E \approx e$ (as flow velocities are typically small
in the heat exchanger channels used for sCO$_2$ cycles~\citep{bone_2018}).
By applying these assumptions to Eqs.\;\ref{sys:q1d_flow_eqs} then
combining Eqs.\;\ref{eq:1DFlowContinuity} and \ref{eq:1DFlowEnergy}, the
energy dynamics of a fluid streams are given by~\cite{bone_2019}
\begin{equation}
\label{eq:e_update_prelim}
\rho\, \pder{e}{t} = - \pder{(\rho h v)}{x} - \frac{q''}{A}, 
\end{equation}
where the pressure and mass flow rate are set by the stream inlet conditions:
$p = p_{in}$ and $\dot{m} = \dot{m}_{in}$.
To obtain a temporal ODE for 
stream internal energy in cell $i$, we integrate Eq.\;\ref{eq:e_update_prelim}
over $N_{cells}$ finite volumes, then apply upwind differencing, giving
\begin{equation}
\label{eq:e_update}
\rho_i\, \der{e_i}{t} = \frac{\dot{m}_{in}\, \left( h_{i-1}\, - h_{i}\right)}{A\, \Delta x}
- \frac{N_{chans}\, \pi\, \text{Nu}\, k_i}{A} \left( T_i\, -    T_{w,\, i} \right).
\end{equation}
%
(See Sec.\;\ref{sec:heat_transfer_correlations} for Nusselt number
correlations.)
For the thermal oil stream, $p_{in}$ and $\dot{m}_{in}$ are set by the pump, and
for the CO$_2$ stream, $p_{in}$ and $\dot{m}_{in}$ are given by the compressor
outflow conditions, which in the control model are captured by the equations
presented in Sec.\;\ref{sec:mass_flow_rate_and_pressure_modelling}.
As for the walls, best closed-loop performance was obtained when retaining only
the primary partial derivatives in the linearization of Eq.\;\ref{eq:e_update}:
\begin{equation}
\begin{split}
\label{eq:e_update_linear}
\rho_{i,\, 0} \frac{de_i}{dt} \ \approx \ &
\frac{\left( h_{i-1,\, 0}\, - h_{i,\, 0}\, \right)}{A\, \Delta x} \overline{\dot{m}}_{in}
\\
                                          & + \frac{\dot{m}_{in,\, 0}}{A\, \Delta x}\,  \left( \pder{h}{e_{i-1,\, 0}}\, \overline{e}_{i-1}\, - \pder{h}{e_{i,\, 0}}\, \overline{e}_{i} \right)
\\
                                          & - \frac{\dot{m}_{in,\, 0}\, \left( h_{i-1,\, 0}\, - h_{i,\, 0}\, \right)\, }{\rho_{i,\, 0}\, A\, \Delta x}\, \pder{\rho}{e_{i,\, 0}}\, \overline{e}_{i}
\\
                                          & - \frac{N_{chans}\, \pi\, \text{Nu}\, k_{i,\, 0}}{\rho_{i,\, 0}\, A}\, \left( \pder{T}{e_{i,\, 0}}\, \overline{e}_{i} - \overline{T}_{w,\, i}\right).
\end{split}
\end{equation}


\subsection{Mass flow rate and pressure modelling}
\label{sec:mass_flow_rate_and_pressure_modelling}

MPC requires a low-order model for the slow-timescale evolution of
mass flow rate and high-side pressure in the system.
This section derives such a model by applying the quasi-steady momentum
assumption, performing system-level mass and momentum balances, and
approximating turbomachinery performance with Taylor series expansions.

First, we perform a system-level momentum balance to relate the turbomachinery
mass flow rates to compressor speed.
For the CO$_2$ stream, the pressure difference between the reservoirs is equal
to the total pressure change over the system:
\begin{equation}
\label{eq:pressure_conservation}
p_{stag,\,in} - p_{stag,\,out} = \Delta p_{pipes} + \Delta p_{HX,\,CO_2} + \Delta p_{c} + \Delta p_{t}
\end{equation}
where the pressure changes over the turbomachinery are
\begin{equation}
\label{eq:delta_p_comp_turb}
\Delta p_{c} = p_{out,\,c} - p_{in,\,c}
\quad \text{and} \quad
\Delta p_{t} = p_{out,\,t} - p_{in,\,t}.
\end{equation}
Assuming the pressure drops over the pipes and heat exchanger ($\Delta
p_{pipes}$ and $\Delta p_{HX,\,CO_2}$) are negligible, the low-side pressures
are fixed to the reservoir pressures
($p_{c,\,in} = p_{stag,\,in}$, $p_{t,\,out} = p_{stag,\,out}$), so
\begin{equation}
\label{eq:pressure_conservation_simplified}
p_{stag,\,in} - p_{stag,\,out} = \Delta p_{c} + \Delta p_{t}
\end{equation}
and
\begin{equation}
\label{eq:high_side_pressure}
p_{high} \equiv p_{c,\,out} \equiv p_{t,\,in}.
\end{equation}

As the low-side pressures are fixed, we reformulate the turbine performance map
model (Eq.\;\ref{eq:performance_map}) with $p_{t,\,out}$ as an input and
$p_{t,\,in}$ as an output, giving
\begin{align}
\Delta p_{c},\, T_{out,\,c} &= g_{comp}(p_{in,\,c},\, T_{in,\,c},\, \dot{m}_{c},\, N_{c})
\\
\Delta p_{t},\, T_{out,\,t} &= g_{turb}(p_{out,\,t},\, T_{in,\,t},\, \dot{m}_{t},\, N_{t}).
\nonumber
\end{align}
Noting that
(1) $p_{in,\,c}$ and $T_{in,\,c}$ are fixed to the inlet reservoir conditions,
(2) $p_{out,\,t}$ is fixed to the outlet reservoir temperature, and
(3) $N_{t}$ is fixed to the synchronous speed,
a first-order Taylor series expansion of $g_{comp}$ and $g_{turb}$ about the current
operating point yields
\begin{align}
\label{eq:delta_p_comp_taylor_simple}
\Delta p_{c} &\approx \Delta p_{c,\,0} + \pder{\Delta p_{c}}{\dot{m}_{c}}\, \overline{\dot{m}}_{c} + \pder{\Delta p_{c}}{N_{c}}\, \overline{N}_{c}
\\
\label{eq:delta_p_turb_taylor_simple}
\Delta p_{t} &\approx \Delta p_{t,\,0} + \pder{\Delta p_{t}}{T_{in,\,t}}\, \overline{T}_{in,\,t}
+ \pder{\Delta p_{t}}{\dot{m}_{t}}\, \overline{\dot{m}}_{t},
\end{align}
where $\pder{\Delta p_{c}}{p_{in,\,c}}$ refers to the partial derivative of
$g_{comp}$ with respect to $p_{in,\,c}$ considering the output variable
$\Delta p_{c}$ (and similar for the other partial derivatives).

Substituting Eqs.\;\ref{eq:delta_p_comp_taylor_simple} and
\ref{eq:delta_p_turb_taylor_simple} into
Eq.\;\ref{eq:pressure_conservation_simplified} yields
\begin{equation}
\begin{split}
\label{eq:pressure_conservation_linearized}
p_{stag,\,in} - p_{stag,\,out} \ = \ &
\Delta p_{c,\,0} + \pder{\Delta p_{c}}{\dot{m}_{c}}\, \overline{\dot{m}}_{c} + \pder{\Delta p_{c}}{N_{c}}\, \overline{N}_{c}
\\
& + \Delta p_{t,\,0} + \pder{\Delta p_{t}}{T_{in,\,t}}\, \overline{T}_{in,\,t} + \pder{\Delta p_{t}}{\dot{m}_{t}}\, \overline{\dot{m}}_{t}.
\end{split}
\end{equation}
As the pressure change between the reservoirs must be equal to the pressure
change over the turbomachinery at the linearization point (i.e.
$p_{stag,\,in} - p_{stag,\,out} = \Delta p_{c,\,0} + \Delta p_{t,\,0}$),
Eq.\;\ref{eq:pressure_conservation_linearized} can be simplified to
\begin{equation}
\label{eq:pressure_conservation_linearized_simplified}
0 = \pder{\Delta p_{c}}{\dot{m}_{c}}\, \overline{\dot{m}}_{c} + \pder{\Delta p_{c}}{N_{c}}\, \overline{N}_{c}
  + \pder{\Delta p_{t}}{T_{in,\,t}}\, \overline{T}_{in,\,t} + \pder{\Delta p_{t}}{\dot{m}_{t}}\, \overline{\dot{m}}_{t}.
\end{equation}
Turbine inlet temperature has only a small effect on high-side pressure, with
the product $\pder{\Delta p_{t}}{T_{in,\,t}}\, \overline{T}_{in,\,t}$ typically
being around three orders of magnitude less than the other terms.
Neglecting this term and differentiating
Eq.\;\ref{eq:pressure_conservation_linearized_simplified} with respect to
time gives the compressor mass flow dynamics as
\begin{equation}
\label{eq:mdot_comp_update}
\der{\dot{m}_{c}}{t} = - \pder{\dot{m}_{c}}{\Delta p_{c}}\, \left( \pder{\Delta p_{c}}{N_{c}}\, \der{N_{c}}{t} + \pder{\Delta p_{t}}{\dot{m}_{t}}\, \der{\dot{m}_{t}}{t} \right),
\end{equation}
where by definition $\der{\overline{N}_{c}}{t} = \der{N_{c}}{t}$
and the same for all other perturbed variables.
Eq.\;\ref{eq:mdot_comp_update} is used to model compressor mass flow dynamics by
substituting the equations for compressor shaft speed dynamics
(Eq.\;\ref{eq:rotordynamics_linearized_simplified}) and turbine mass flow
dynamics (Eq.\;\ref{eq:mdot_turb_update}).

Next, we develop an equation for turbine mass flow rate by applying the
conservation of mass to the high-pressure side of the system. 
The fluid mass in the high-pressure side is
\begin{equation}
\label{eq:high_side_mass}
m_{high} = \sum_{i=1}^{N_{cells,\,high}} \rho_i\, V_i,
\end{equation}
where $N_{cells,\,high}$ is the number of cells in the high-pressure side.
Additionally, the rate that fluid mass accumulates in the high-pressure side of
the system is the difference between the compressor and turbine mass flow rates:
\begin{equation}
\label{eq:conservation_of_mass}
\der{m_{high}}{t} = \dot{m}_{c} - \dot{m}_{t}.
\end{equation}
As cell volume is constant, differentiating Eq.\;\ref{eq:high_side_mass} with
respect to time yields
\begin{equation}
\label{eq:dm_dt_1}
\der{m_{high}}{t} = \sum_{i=1}^{N_{cells,\,high}} \der{\rho_i}{t}\, V_i.
\end{equation}
Choosing thermodynamic state variables of $\rho$
and $e$, Eq.\;\ref{eq:dm_dt_1} can be expanded as
\begin{equation}
\label{eq:dmhigh_dt}
\der{m_{high}}{t} = \sum_{i=1}^{N_{cells,\,high}} V_i
\left( \pderv{\rho}{e}{\begin{smallmatrix}p = p_i \\e = e_i \end{smallmatrix}}\, \der{e_i}{t} + \pderv{\rho}{p}{\begin{smallmatrix}p = p_i \\e = e_i \end{smallmatrix}}\, \der{p_i}{t} \right),
\end{equation}
where $p_{i,\,0}$ and $e_{i,\,0}$ are the pressures and internal energies in
cell $i$ at the current operating point.
Because we assume the high-side pressure to be uniform, Eq.\;\ref{eq:dmhigh_dt}
can be simplified as
\begin{equation}
\begin{split}
\label{eq:dm_high_dt}
\der{m_{high}}{t} \ = \ & \left( \sum_{i=1}^{N_{cells,\,high}} V_i\, \pderv{\rho}{e}{\begin{smallmatrix*}[l]p = p_{high} \\e = e_i \end{smallmatrix*}}\, \der{e_i}{t} \right) \;
\\
                        & + \; \left( \sum_{i=1}^{N_{cells,\,high}} V_i\, \pderv{\rho}{p}{\begin{smallmatrix*}[l]p = p_{high} \\e = e_i \end{smallmatrix*}} \right)\, \der{p_{high}}{t}.
\end{split}
\end{equation}
From Eqs.\;\ref{eq:delta_p_comp_turb} and \ref{eq:high_side_pressure}, the
time derivatives of $\Delta p_{t}$ and $p_{high}$ are related as follows:
\begin{equation}
\der{}{t} \Delta p_{t} = \der{}{t} (p_{out,\,t} - p_{in,\,t}), 
\end{equation}
so
\begin{equation}
\label{eq:dp_high_to_ddelta_pt}
\der{p_{high}}{t} = -\der{\Delta p_{t}}{t}.
\end{equation}
Deriving the Taylor series approximation for $\Delta p_{t}$
(Eq.\;\ref{eq:delta_p_turb_taylor_simple}) with respect to time and
substituting into Eq.\;\ref{eq:dp_high_to_ddelta_pt}, gives
\begin{equation}
\label{eq:dp_high_dt_prelim}
\der{p_{high}}{t} = - \pder{\Delta p_{t}}{T_{in,\,t}}\, \der{T_{in,\,t}}{t} - \pder{\Delta p_{t}}{\dot{m}_{t}}\, \der{\dot{m}_{t}}{t}.
\end{equation}
%
As discussed, turbine pressure ratio is not strongly influenced by inlet
pressure, so we assume
\begin{equation}
\label{eq:p_high_update}
\der{p_{high}}{t} = - \pder{\Delta p_{t}}{\dot{m}_{t}}\, \der{\dot{m}_{t}}{t}.
\end{equation}
%
Combining Eqs.\;\ref{eq:dm_high_dt} and \ref{eq:p_high_update},
substituting into Eq.\;\ref{eq:conservation_of_mass}, then rearranging gives the
turbine mass flow dynamics as
\begin{equation}
\begin{split}
\label{eq:mdot_turb_update}
\der{\dot{m}_{t}}{t} \ = \ & \left( \pder{\Delta p_{t}}{\dot{m}_{t}}\, \sum_{i=1}^{N_{cells,\,high}} V_i\, \pderv{\rho}{p}{\begin{smallmatrix*}[l]p = p_{high} \\e = e_i \end{smallmatrix*}} \right)^{-1}\,
\\
                           & \times \left( \dot{m}_{t} - \dot{m}_{c} +  \sum_{i=1}^{N_{cells,\,high}} V_i\, \pderv{\rho}{e}{\begin{smallmatrix*}[l]p = p_{high} \\e = e_i \end{smallmatrix*}}\, \der{e_i}{t} \right).
\end{split}
\end{equation}
Due to the damping effect of component wall thermal inertia, the dynamics of
internal energy are slower than those for mass flow rate and pressure.
Accordingly, we model turbine mass flow dynamics by substituting the current
values of internal energy time derivatives $\der{e_{i}}{t}_0$ into
Eq.\;\ref{eq:mdot_turb_update}.

This concludes development of the control dynamics model.
This model comprises equations for the dynamics of
wall temperature (Eq.\;\ref{eq:wall_temp_update}),
fluid stream internal energy (Eq.\;\ref{eq:e_update}),
mass flow rates and pressure in the CO$_2$ stream
(Eqs.\;\ref{eq:mdot_comp_update}, \ref{eq:mdot_turb_update},
\ref{eq:p_high_update}),
compressor shaft speed
(Eq.\;\ref{eq:rotordynamics_linearized}), and thermal oil pump system
(Eq.\;\ref{eq:pump_dynamics}).


\subsection{Output model --- net power and turbine inlet temperature}
\label{sec:output_model_net_power_and_turbine_inlet_temperature}

For the control problem defined in Sec.\;\ref{sec:control_problem}, the
tracked-outputs are net power output and turbine inlet temperature.
Choosing thermodynamic state variables of $p$ and $e$, turbine inlet temperature
can be approximated from $p_{high}$ and the final-cell high-side internal energy
$e_{N_{cells,high}}$ as
\begin{equation}
\label{eq:T_t_in_state_vars}
\overline{T}_{in,\,t} = T_{in,\,t,\,0}
+ \pderv{T}{e}{\begin{smallmatrix*}[l]p = p_{high} \\ e = e_{in,\,t} \end{smallmatrix*}}\, \overline{e}_{in,\,t}
+ \pderv{T}{p}{\begin{smallmatrix*}[l]p = p_{high} \\ e = e_{in,\,t} \end{smallmatrix*}}\, \overline{p}_{high},
\end{equation}
where $T_{in,\,t,\,0}$ is the most recent measurement of turbine inlet temperature.
For the open sCO$_2$ cycle, the net power output is
\begin{equation}
\label{eq:p_net}
\dot{W}_{net} = \dot{W}_{t} - \dot{W}_{c}.
\end{equation}
Linearizing the performance maps and discarding negligible terms (as for $\Delta
p_c$ and $\Delta p_t$) yields
\begin{equation}
\begin{split}
\label{eq:p_net_control}
\dot{W}_{net} \ = \ &
                    \dot{W}_{t,\,0} \;-\; \dot{W}_{c,\,0}
                    \;+\; \pder{\dot{W}_{t}}{T_{in,\,t}}\, \overline{T}_{in,\,t}
                    \;+\; \pder{\dot{W}_{t}}{\dot{m}_{t}}\, \overline{\dot{m}}_{t}
\\
                    & \;-\; \pder{\dot{W}_{c}}{\dot{m}_{c}}\, \overline{\dot{m}}_{c}
                    \;-\; \pder{\dot{W}_{c}}{N_{c}}\, \overline{N}_{c}.
\end{split}
\end{equation}
Turbine inlet temperature has a significant effect on shaft power and so must be
retained in Eq.\;~\ref{eq:p_net_control}.


\subsection{Compressor speed constraints}
\label{sec:compressor_speed_constraints}


The compressor is subject to minimum and maximum speed constraints.
The maximum speed $N_{c,\,max}$ is dictated by the shaft bearings and is taken
as 126\% of the nominal compressor speed based on manufacturer data.
Applying some constraint margin $k_{N,\,max}$, the maximum compressor speed is
\begin{equation}
N_c \ \leq \ (1 - k_{N,\,max})\, N_{c,\,max}.
\end{equation}

Minimum compressor speed is dictated by surge: a damaging physical phenomenon,
characterized by oscillatory or reversed flow, which occurs when the compressor
cannot achieve the required pressure ratio.
This section shows how to compute the surge conditions for a given pair of
turbomachines.
We consider only the case where the turbine operates at a fixed synchronous
speed $N_{t0}$.

\begin{figure}[h]
\centering
\includegraphics[width=0.65\columnwidth]{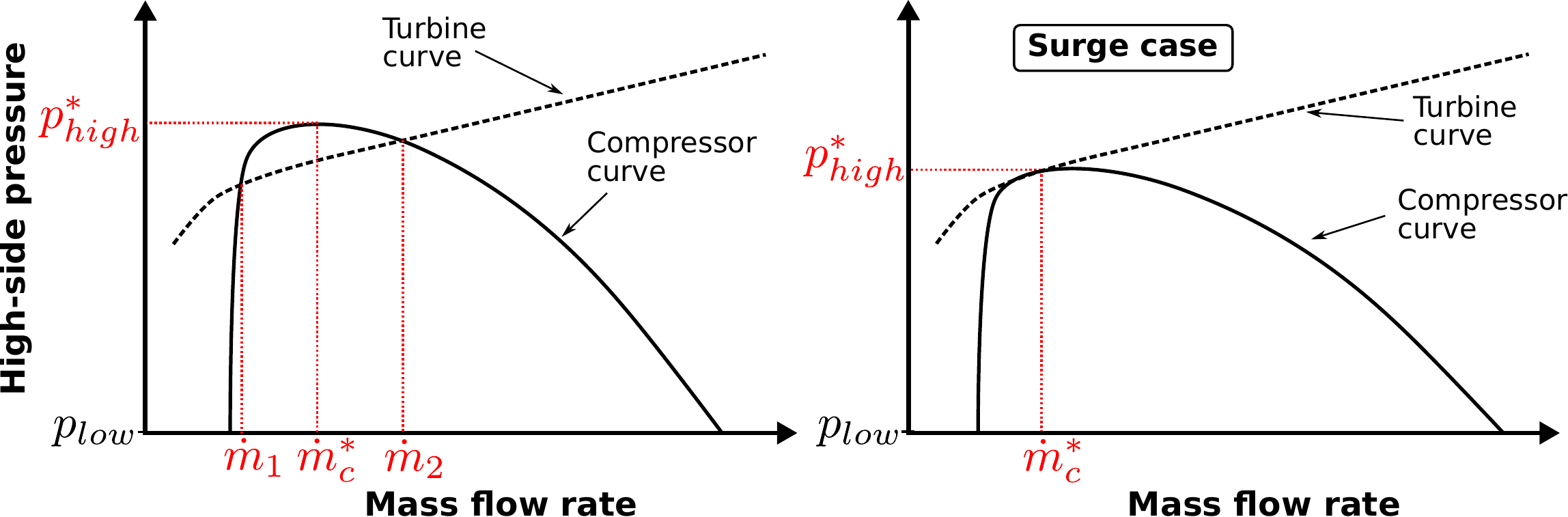}
    
    
\caption{
High-side pressure versus mass flow curves for both turbomachines.
}
\label{fig:p_high_vs_mdot}
\vspace{-10pt}
\end{figure}

In the control model, the compressor and turbine both have the same high-side
pressure $p_{high} \equiv p_{out,\,c} \equiv p_{in,\,t}$, and at steady-state,
they both have the same mass flow rate
$\dot{m}_{sys} \equiv \dot{m}_c \equiv \dot{m}_t$.
The low-side pressures are fixed, so for given shaft speeds and inlet
temperatures, the feasible system mass flow rates are given by the intersections
of the $\dot{m}$ versus $p_{high}$ curves for the two turbomachines.
Fig.\;\ref{fig:p_high_vs_mdot} shows some sample $\dot{m}$ versus $p_{high}$
curves, with the feasible mass flow rates $\dot{m}_1$ and $\dot{m}_2$ marked.
As we employ surge control with a margin, the system is forced to operate at
$\dot{m}_2$.


Surge conditions can be computed by considering the effect of varying
compressor speed.
As compressor speed decreases, the maximum achievable high-side pressure
$p_{high}^*$ drops.
Surge occurs at the minimum feasible compressor speed $N_{c,\,min}$, where the
compressor and turbine $p_{high}$ versus $\dot{m}$ curves intersect only at
$p_{high}^*$ (as illustrated on the right side of
Fig.\;\ref{fig:p_high_vs_mdot}).
Below this speed, the compressor cannot supply the high-side pressure required
by the turbine at any mass flow rate, causing surge.


$N_{c,\,min}$ is operating-point dependent and can be computed online using the
following procedure.
For the chosen compressor maps (see~\ref{sec:turbine_map_model}), 
we store the combinations of mass flow rate and corrected speed that
maximize the compressor outlet pressure, then store these values in lookup
tables
$p_{high}^* = f_{c1}(N_{cor})$
and
$\dot{m}_c^* = f_{c2}(N_{cor})$.
For the chosen turbine map model, the high-side pressure $p_{high} = p_{in,\,t}$
can be computed from Eq.\;\ref{eq:turb_map_1} as
\begin{equation} \label{eq:p_high_star_turb} p_{high} = p_{out,\,t}\;
M_{t1}\left(\dot{m}_{t}\, \sqrt{T_{in,\,t}} / p_{in,\,t}\right).
\end{equation}
Assuming that the current difference between the turbine and compressor mass
flow rates $\Delta \dot{m}$ persists, the mass flow rate at the turbine when the
compressor achieves $p_{high}^*$ is $\dot{m}_t^* = \dot{m}_c^* + \Delta \dot{m}$.
%
Surge occurs when the maximum compressor outlet pressure equals the turbine
inlet pressure ($p_{high}^* = p_{high}$):
%
\begin{equation}
\label{eq:N_c_min_prelim}
f_{c1}(N_{cor}) = p_{out,\,t}\; M_{t1}\left(\dot{m}_{t}^*\, \sqrt{T_{in,\,t}} /
p_{in,\,t}\right)
\end{equation}
To solve for $N_{c,\,min}$, we note that $p_{in,\,t} = p_{high}^*$, substitute
Eq.\;\ref{eq:mc_star} into
Eq.\;\ref{eq:N_c_min_prelim}, then reformulate to give
\begin{equation}
\label{eq:N_c_min_root_finding}
0 = f_{c1}(N_{cor}) - p_{out,\,t}\; M_{t1}\left( \frac{ \sqrt{T_{in,\,t}} }{ f_{c1}(N_{cor}) } (f_{c2}(N_{cor}) + \Delta \dot{m}) \right).
\end{equation}
We compute $N_{cor,\,min}$ for the current values of $p_{out,\,t}$, $\Delta
\dot{m}$, and $T_{in,\,t}$ by solving the root-finding problem defined by
Eq.\;\ref{eq:N_c_min_root_finding}, then compute the actual minimum speed
$N_{c,\,min}$ using Eq.\;\ref{eq:n_s_corr}.

To prevent surge, we impose the minimum speed constraint
\begin{equation}
N_c \ \geq \ (1 + k_{N,\,min}) N_{c,\,min,\,0}
\end{equation}
where $N_{c,\,min,\,0}$ is the minimum speed for the current operating point,
$k_{N,\,min}$ is the constraint margin, and
the partial derivatives are computed using finite differences from
Eq.\;\ref{eq:N_c_min_root_finding}.
Again, turbine inlet temperature is approximated using
Eq.\;\ref{eq:T_t_in_state_vars}.
We set $k_{N,\,min} = 5$\% and $k_{N,\,max} = 5$\%.



\section{Controller design}
\label{sec:controller_design}

This section presents an MPC scheme that regulates the net power
output $\dot{W}_{net}$ of the open sCO$_2$ cycle (see
Sec.\;\ref{sec:control_problem}) to $\dot{W}_{net,\,ref}$, while driving the
turbine inlet temperature to the target point $T_{in,\,t,\,ref}$.
To maximize the system's thermodynamic efficiency, $T_{in,\,t,\,ref}$ is set as
the maximum feasible turbine inlet temperature for the net power
$\dot{W}_{net,\,ref}$.
The control inputs are the reference flow rate of the thermal oil pump
$\dot{m}_{oil,\,ref}$ and the compressor motor torque $T_m$, which are
parametrized using zeroth-order and first-order holds respectively.


\subsection{State-space form}
\label{sec:state_space_form}

To implement MPC, we cast the control model in explicit state-space form:
\begin{align}
\label{eq:state_space_form}
\frac{d\mbf{x}}{dt} & = f_c(\mbf{x},\, \mbf{u}) \\
            \mbf{z} & = g_c(\mbf{x},\, \mbf{u}) \nonumber
\end{align}
where $\mbf{x} \in \mathbb{R}^n$, $\mbf{u} \in \mathbb{R}^m$, and $\mbf{z} \in
\mathbb{R}^l$ are the state, input, and tracked-output vectors respectively, and
$f_c: \mathbb{R}^{n} \times \mathbb{R}^{m} \to \mathbb{R}^n$ and $g_c:
\mathbb{R}^{n} \times \mathbb{R}^{m} \to
\mathbb{R}^l$ represent the nonlinear dynamics and output control models
respectively.
%
The input, tracked-output, and reference vectors are
\begin{align}
\mbf{u} &= \left[ \dot{T}_{m},\, \dot{m}_{oil,ref} \right]^\intercal
\\
\mbf{z} &= \left[ \dot{W}_{net},\, T_{in,\,t} \right]^\intercal \nonumber
\\
\mbf{r} &= \left[ \dot{W}_{net,\,ref},\, T_{in,\,t,\,ref} \right]^\intercal. \nonumber
\end{align}
$f_c$ comprises equations for the dynamics of internal energy in the fluid streams
(Eq.\;\ref{eq:e_update}), wall temperature
(Eq.\;\ref{eq:wall_temp_update}), mass flow rates and pressure in the CO$_2$
stream (Eqs.\;\ref{eq:mdot_comp_update}, \ref{eq:mdot_turb_update}, and
\ref{eq:p_high_update}), the compressor shaft speed
(Eq.\;\ref{eq:rotordynamics_linearized}), and the thermal oil pump system
(Eq.\;\ref{eq:pump_dynamics}).
We choose thermodynamic state variables of density and pressure (and compute all
other thermodynamic properties using the equation of state), so the state
vector is
\begin{equation}
\mbf{x} = \left[ \mbf{T}_{wall}^\intercal,\, \mbf{e}_{CO_2}^\intercal,\, p_{high},\, \dot{m}_c,\, \dot{m}_t,\, N_{c},\, \mbf{e}_{oil}^\intercal,\, \dot{m}_{oil} \right]^\intercal,
\end{equation}
%
where $\mbf{T}_{wall} = [ T_{wall,\,1},\, ... ,\, T_{wall,\,N_{cells,\,HX}} ]^\intercal$,
$\mbf{e}_{CO_2} = [ e_{CO_2,\,1},\, ... ,\, e_{CO_2,\,N_{cells}} ]^\intercal$, and
$\mbf{e}_{oil} = [ e_{oil,\,1},\, ... ,\, e_{oil,\,N_{cells,\,HX}} ]^\intercal$.
$g_c$ comprises the equations for $T_{in,\,t}$ and $\dot{W}_{net}$
(Eqs.\;\ref{eq:T_t_in_state_vars} and \ref{eq:p_net_control}).


\subsection{Linearization and discrete-time conversion}
\label{sec:linearization_and_discrete-time_conversion}


We linearize the control model online about the current operating point
($\mbf{x}_0$, $\mbf{u}_0$) so that the control updates can be computed using a
quadratic programming (QP) solver.
Using perturbed variables $\mbf{x}' = \mbf{x} - \mbf{x}_0$ and
$\mbf{u}' = \mbf{u} - \mbf{u}_0$, a local linear time-invariant model is
given by
\begin{align}
\label{eq:continuous_lti_model}
\frac{d\mbf{x}}{dt} & = A\, \mbf{x}' + B\, \mbf{u}' + \mbf{f}_0
\\
            \mbf{z} & = C\, \mbf{x}' + D\, \mbf{u}' + \mbf{g}_{0}, \nonumber
\end{align}
where $A$, $B$, $C$, $D$ represent the Jacobian matrices of $f_c$ and $g_c$ with
respect to $\mbf{x}$ and $\mbf{u}$, $\mbf{f_{0}} = f_c(\mbf{x}_0,
\mbf{u}_0)$, and $\mbf{g_{0}} = g_c(\mbf{x}_0, \mbf{u}_0)$.
$D$ is always zero and thus neglected from now on.
Only certain terms are retained in the linearized dynamics equations for heat
exchanger wall temperature stream internal energy and (see
Eqs.\;\ref{eq:wall_temp_update_linear} and \ref{eq:e_update_linear}).
We implement MPC with a discrete-time version of
Eq.\;\ref{eq:continuous_lti_model} with sampling time $\Delta t$,
\begin{align}
\label{eq:discrete_lti_model}
\mbf{x}'_{k+1} &= A_d\, \mbf{x}'_{k} + B_d\, \mbf{u}'_{k} + \mbf{f}_{0,\,d}
\\
\mbf{z}_{k}    &= C\, \mbf{x}'_{k} + \mbf{g}_{0}, \nonumber
\end{align}
where~\citep{franklin_1990_ch2}
\begin{align}
\label{eq:discrete_time_conversion}
A_d   &= \exp{(A\, \Delta t)} \nonumber
\\
\Gamma &= \int^{\Delta t}_0{\exp{(A\, (\Delta t - \tau))}}\, d\tau\ I
\\
B_d    &= \Gamma\, B, \quad \mbf{f}_{0,d} = \Gamma\, \mbf{f}_0 \nonumber
\end{align}
The matrix exponentials are computed using the scaling and squaring
method~\citep{higham_2005}.
In the controller, Eq.~\ref{eq:discrete_lti_model} is augmented with an output
disturbance model for offset-free reference tracking (see
Eq.~\ref{eq:tracked_outputs_augmented}).


\subsection{State estimation}
\label{sec:state_estimation}

To implement MPC, we require an estimate $\hat{\mbf{x}}$ of the state vector
$\mbf{x}$.
We assume that the mass flow rates, turbomachinery speeds, and high-side
pressure are directly measureable.
However, the states inside the heat exchanger (the internal energies and wall
temperatures) are not measureable, so we estimate these states
\begin{equation}
\mbf{x}_{e} = \left[ \mbf{T}_{wall}^\intercal,\, \mbf{e}_{CO_2}^\intercal, \mbf{e}_{oil}^\intercal \right]^\intercal
\end{equation}
with an observer using measurements of the inlet and outlet fluid temperatures
of each stream
\begin{equation}
\mbf{y} = \left[ T_{CO_2,\,in},\, T_{CO_2,\,out},\, T_{oil,\,in},\, T_{oil,\,out} \right]^\intercal,
\end{equation}
where $\mbf{y} \in \mathbb{R}^l_y$.
We estimate $\mbf{x}_{e}$ using a reduced submodel that describes the dynamics
of wall temperature (Eq.\;\ref{eq:wall_temp_update}) and fluid stream internal
energy (Eq.\;\ref{eq:e_update}), treating the mass flow rates as inputs.
In this model, the output temperatures in $\mbf{y}$ are computed analogously to
Eq.~\ref{eq:T_t_in_state_vars}.
The discrete-time linearization of this observer submodel is represented by
matrices $A_e$, $B_e$, $C_e$ and vectors $\mbf{x}'_e$, $\mbf{u}'_e$,
$\mbf{f}_{0,\,e}$ and $\mbf{g}_{0,\,e}$ which are computed analogously to those
used for control (see Sec.~\ref{sec:linearization_and_discrete-time_conversion}).

We use an extended Kalman filter (EKF)~\citep{gelb_1974} to compute the state
estimate $\hat{\mbf{x}}$ from the measurements $\mbf{y}$.
Due to the approximations and coarser discretization used in the control model,
there is significant mismatch between the control and simulation models.
To achieve offset-free reference tracking under this plant-model mismatch, we
augment the control model with an output disturbance $\mbf{d} \in
\mathbb{R}^{l_y}$ acting on each measurement~\citep{maeder_2009}.
A similar approach would be required for a real plant, where the true behaviour
can never be modelled exactly.

The dynamics and outputs of this augmented system are~\citep{maeder_2009}
\begin{equation}
\label{eq:state_space_dynamics_augmented_linear}
\begin{bmatrix} \mbf{x}'_{e,\,k+1} \\ \mbf{d}_{k+1} \end{bmatrix}
  =
\begin{bmatrix} A_e & 0  \\ 0 & I \end{bmatrix}
\begin{bmatrix} \mbf{x}'_{e,\,k} \\ \mbf{d}_k \end{bmatrix}
 +
\begin{bmatrix} B_e \\ 0 \end{bmatrix} \mbf{u}'_{e,\,k}
 +
\begin{bmatrix} \mbf{f}_{0,\,e} \\ 0 \end{bmatrix}
\end{equation}
and
\begin{equation}
\label{eq:state_space_outputs_augmented_linear}
\mbf{y}_k = \begin{bmatrix} C_y & I \end{bmatrix} \begin{bmatrix} \mbf{x}'_{e,\,k} \\ \mbf{d} \end{bmatrix} + \mbf{g}_{0,\,e}.
\end{equation}
Defining the augmented state vector $\mbf{\chi}_k = [\mbf{x}'^\intercal_{e,\,k}\;
\mbf{d}^\intercal_k]^\intercal$, we write the model given by
Eqs.~\ref{eq:state_space_dynamics_augmented_linear} and
\ref{eq:state_space_outputs_augmented_linear} as 
%
\begin{align}
\label{eq:AugmentedXi}
\mbf{\chi}_{k+1} & = A_{a}\, \mbf{\chi}_{k} + B_{a}\, \mbf{u}'_{k} + \mbf{f}_{0\,{a}}
\\
   \mbf{y}_{k} & = C_{a}\, \mbf{\chi}_{k} + \mbf{g}_{0,\,e}. \nonumber
\end{align}

The EKF assumes that the augmented system dynamics are given by
\begin{align}
\label{eq:EkfDynamics}
\mbf{\chi}_{k+1} &= F(\mbf{\chi}_{k},\, \mbf{u}'_{k}) + \mbf{w}_k
\\
    \mbf{y}_{k} &= G(\mbf{\chi}_{k},\, \mbf{u}'_{k}) + \mbf{v}_k, \nonumber
\end{align}
where
$F$ and $G$ represent the dynamics and output models in
Equation~\ref{eq:AugmentedXi}, and
$\mbf{w}_k$ and $\mbf{v}_k$ are zero-mean white noise processes with covariance matrices $\Sigma_w
\in \mathbb{R}^{n+l_y\times n+l_y}$ and $\Sigma_v \in \mathbb{R}^{{l_y}\times {l_y}}$
respectively.
Furthermore, it assumes that at timestep $k$, the state estimate 
$\hat{\mbf{\chi}}_{k}$ is a normally-distributed random variable with mean
$\hat{\bar{\mbf{\chi}}}_{k}$ and covariance $\Sigma_{x,\,k} \in \mathbb{R}^{n_\chi\times
n_\chi}$.
We denote the mean of the output estimate corresponding to $\hat{\mbf{\chi}}_{k}$
as $\hat{\bar{\mbf{y}}}_{k}$.

We update the state estimate by first computing the predicted next-timestep
state and output estimates as
\begin{align}
\label{eq:EkfPredict}
\bar{\hat{\mbf{\chi}}}_{k}^\dagger &= F(\bar{\hat{\mbf{\chi}}}_{k-1},\, \bar{\mbf{u}}_{k-1})
\\
\bar{\hat{\mbf{y}}}_{k}^\dagger &= G(\bar{\hat{\mbf{\chi}}}_{k}^\dagger) \nonumber
\\
                    \Sigma_{x,\,k}^\dagger &=  A_a\, \Sigma_{x,\,k-1}\, {A_a}^\intercal + \Sigma_w. \nonumber
\end{align}
Then, using the measurements $\mbf{y}_{k}$, we correct the state estimate
according to 
\begin{align}
\label{eq:EkfCorrect}
            \mbf{i}_{k} &= \mbf{y}_{k} - \bar{\hat{\mbf{y}}}_{k}^\dagger
\\
                    K_k &= \Sigma_{x,\,k}^\dagger\, {C_a}^\intercal\, \left( C_a\, \Sigma_{x,\,k}^\dagger\, {C_a}^\intercal + \Sigma_v\right)^{-1} \nonumber
\\
\bar{\hat{\mbf{\chi}}}_{k} &= \bar{\hat{\mbf{\chi}}}_{k}^\dagger \;+\; K_k\, \mbf{i}_{k} \nonumber
\\
                      \Sigma_{x,\,k} &= (I - K_k\, C_a)\, \Sigma_{x,\,k}^\dagger\, (I - K_k\, C_a)^\intercal\, + K_k\, \Sigma_{v,\,k}\, K_k^\intercal, \nonumber
\end{align}
where $\mbf{i}_k$ is the innovation and $K_k$ is the Kalman gain.

We set values for the noise covariance matrices based on nominal values for the
estimated states and outputs ($T_{CO_2,\,nom}$, $e_{CO_2,\,nom}$,
$T_{oil,\,nom}$, $e_{oil,\,nom}$ and $T_{wall,\,nom}$ ---
see Tab.~\ref{tab:nominal_vars} for values).
We assume that all measurements are independent and that measurement error is
normally distributed with zero mean and variance of 2\% of the nominal measured
value.
Thus, the measurement noise covariance is
\begin{equation}
\Sigma_v = 0.02\, \times\, \text{diag}\begin{pmatrix}  T_{CO_2,\,nom} & T_{CO_2,\,nom} & T_{oil,\,nom} & T_{oil,\,nom} \end{pmatrix}.
\end{equation}
We assume that the continuous-time process noise $\Sigma_{w,\,c}$ is diagonal with a power
spectral density of one.
Additionally, we assume that variance of the state variables ($\mbf{e}_{hot}$,
$\mbf{e}_{cold}$, and $\mbf{T}_{wall}$) is 10\% of nominal and that the variance
of the disturbances variables is 5\% of nominal.
Under these assumptions, the continuous-time process noise is
\begin{equation}
\begin{split}
\Sigma_{w,\,c} = \text{diag}\left( 0.1\,\mbf{T}_{wall,\,nom} \enskip 0.1\,\mbf{e}_{CO_2,\,nom}  \enskip 0.1\,\mbf{e}_{oil,\,nom}\right.
\\
\left.
0.05\,T_{CO_2,\,nom} \enskip 0.05\,T_{CO_2,\,nom} \enskip 0.05\,T_{oil,\,nom} \enskip 0.05\,T_{oil,\,nom} \right),
\end{split}
\end{equation}
where $\mbf{e}_{CO_2,\,nom}$ is a vector of length $N_{cells}$
($\mbf{e}_{CO_2,\,nom} = \left[e_{CO_2,\,nom},\, ...,\, e_{CO_2,\,nom}\right]$)
and similar for $\mbf{e}_{cold,\,nom}$ and $\mbf{T}_{wall,\,nom}$.
Using the most recent augmented LTI model, we approximate the discrete-time
process noise as
\begin{equation}
\Sigma_w = A_a\, \Sigma_{w,\,c}\, A_a^\intercal\, \Delta t.
\end{equation}
%
%
Our state estimation approach accounts for model mismatch in the models of the
fluid streams, but not in the turbomachinery performance maps.
This approach is reasonable, since for a real plant, the performance maps could
be frequently updated based on real operating data, removing mismatch.
If accurate performance maps were not used, the observer would need to also
estimate output power (with an associated disturbance) to achieve offset-free
tracking.

\begin{table}[!h]
\renewcommand{\arraystretch}{0.9}
\linespread{1.0}\selectfont\centering
{\footnotesize \caption{Nominal variable values for estimator tuning\label{tab:nominal_vars}}
\begin{tabular}{lrr}
\toprule
\textbf{Variable}       & \textbf{Value}                          \\
\midrule                                           
$p_{oil,\,nom}$:        & \SI{4.0}{MPa}                           \\
$T_{oil,\,nom}$:        & \SI{570}{K}                             \\
$e_{oil,\,nom}$:        & EOS($p_{oil,\,nom}$, $T_{oil,\,nom}$)   \\
$p_{CO_2,\,nom}$:       & \SI{12.0}{MPa}                          \\
$T_{CO_2,\,nom}$:       & \SI{520}{K}                             \\
$e_{CO_2,\,nom}$:       & EOS($p_{CO_2,\,nom}$, $T_{CO_2,\,nom}$) \\
\bottomrule
\end{tabular}
}
\end{table}


\subsection{MPC formulation}
\label{sec:mpc_formulation}

Here, we develop a model predictive controller for the control problem presented
in Sec.~\ref{sec:control_problem}.
We use tracking MPC~\citep{maciejowski_2002} with the estimator detailed in
Sec.~\ref{sec:state_estimation}.
We compute the tracked outputs as
\begin{equation}
\label{eq:tracked_outputs_augmented}
\mbf{z}_k = C_z\, \mbf{x}'_k + C_d\, \mbf{d}_k + \mbf{g}_{0}
\end{equation}
where the matrix $C_d$ maps the disturbances on to the tracked outputs.
Since $T_{CO_2,\,out} = T_{in,\,t}$,
\begin{equation}
C_d = \begin{pmatrix} 0 & 0 & 0 & 0 \\ 0 & 1 & 0 & 0 \end{pmatrix}.
\end{equation}
From the current reference vector $\mbf{r}_k$, we compute the target state $\mbf{x}'_r$
and input $\mbf{u}'_r$ by solving the continuous-time system
\begin{equation}
\label{eq:terminal_state}
\begin{bmatrix} A & B \\ C_z & 0 \end{bmatrix}
\begin{bmatrix} \mbf{x}'_r \\ \mbf{u}'_r \end{bmatrix}
=
\begin{bmatrix}
-\mbf{f}_0 \\ \mbf{r}_k - C_d\, \hat{\mbf{d}}_k - \mbf{g}_{0}
\end{bmatrix}
\end{equation}
for the current disturbance estimate $\hat{\mbf{d}}_k$ (see Sec.~\ref{sec:state_estimation}).
We use tracking MPC with a terminal state cost and costs on control moves, so
the cost functional takes the form~\citep{maciejowski_2002}
\begin{equation}
\label{eq:cost_functional}
J(k,\, \mbf{x}_k,\, \mbf{U}) =
      \left\lVert \mbf{x}'_{k+H_p} - \mbf{x}'_r \right\rVert^2_{P}
\;+\; \sum_{i = 0}^{H_p-1} \left\lVert \mbf{x}'_{k+i} - \mbf{x}'_r \right\rVert^2_Q + \left\lVert \Delta \mbf{u}_{k+i} \right\rVert^2_R,
\end{equation}
where $H_p$ is the prediction horizon, $\mbf{U}$ is the vector of control inputs
\begin{equation}
\mbf{U} = \left[\mbf{u}_k'^\intercal,\, \mbf{u}_{k+1}'^\intercal,\, ...,\, \mbf{u}_{k+H_p-1}'^\intercal\right]^\intercal,
\end{equation}
$\mbf{\Delta u}_{k+i} = \mbf{u}'_{k+i} - \mbf{u}'_{k+i-1}$, and $Q$, $P$, and $R$ are
stage, terminal, and input costs.
We compute the stage cost from the output cost $Q_z \in \mathbb{R}^{l\times l}$
as $Q = C_z^\intercal Q_z C_z$, so that
\begin{equation}
\label{eq:stage_output_cost_equivalance}
\left\lVert \mbf{x}' - \mbf{x}'_r \right\rVert^2_{Q} \equiv \left\lVert \mbf{z} - \mbf{r} \right\rVert^2_{Q_z}.
\end{equation}
As the plant is open-loop stable, we can obtain a stable control law by
computing the terminal cost as the solution of the discrete-time matrix Lyapunov
equation~\citep{maciejowski_2002}
\begin{equation}
\label{eq:matrix_lyapunov_equation}
A_d^\intercal P A_d - P + Q = 0.
\end{equation}
At each $k$, to compute the next control update $\mbf{u}_k$, we solve the
constrained QP
\begin{equation}
\label{eq:mpc_qp}
\begin{aligned}
& \min_{\mathbf{U}} & & J(k,\, \mbf{x}_k,\, \mbf{U})
\\
& \text{subject to} & & \mbf{x}'_{k+i+1} = A_d\, \mbf{x}'_{k+i} + B_d\, \mbf{u}'_{k+i} + \mbf{f}_{0,\,d}
\\
&                   & & \quad \text{for} \enskip i = 1,\, ...,\, H_p - 1
\\
&                   & & \mbf{z} \in \mathcal{Z}
\\
&                   & & \mbf{u} \in \mathcal{U}
\end{aligned}
\end{equation}
for the current augmented state estimate ($\hat{\mbf{x}}_k$, $\hat{\mbf{d}}_k$)
and previous control input $\mbf{u}_{k-1}$.
$\mathcal{U}$ and $\mathcal{Z}$ represent the admissible input and output sets,
which are discussed in Sec.\;\ref{sec:constraints}.
Our model has a relatively high state dimensionality due to the discretization
of the fluid streams.
Thus, we formulate the QP using dense matrices (as in \citet{maciejowski_2002}),
so its computational complexity depends only on $m$ and $l$ (and not
$n$)~\citep{wang_2010}.
We solve the QPs using \texttt{Gurobi}~\citep{gurobi}.
Using slack variables, we implement the constraints on $\mbf{z}$ as soft
constraints with weightings discussed in Sec.\;\ref{sec:constraints}.
The constraint weights are set such that constraint violations are negligible
for all tested cases.

%


\subsection{Constraints}
\label{sec:constraints}

In addition to the compressor speed constraints (see
Sec.\;\ref{sec:compressor_speed_constraints}), we also consider input
constraints and a maximum turbine inlet temperature constraint.
To guarantee feasibility of the optimization problems, state and output
constraints are implemented as soft constraints using weighted linear
penalization (see Tab.\;\ref{tab:constraints} for parameter values).
Constraint weights were set such that state and output constraint violations
are negligible.

\begin{table}[!h]
\renewcommand{\arraystretch}{0.9}
\linespread{1.0}\selectfont\centering
{\footnotesize \caption{Constraint bounds and soft constraint parameters\label{tab:constraints}}
\begin{tabular}{lrrr}
\toprule
\textbf{Variable}            & \textbf{Constraint}                                & \textbf{Weight} \\
\midrule
Compressor speed:            & See Sec.\;\ref{sec:compressor_speed_constraints}   & 1E6             \\
Turbine inlet temperature:   & $T_{in,\,t} \leq $ \SI{570}{K}                     & 2000            \\
Motor torque:                & $0 \leq T_{motor} \leq $ \SI{200}{Nm}              & -               \\
                             & $-15 \leq \dot{T}_{motor} \leq $ \SI{15}{Nm/s}     & -               \\
Heat transfer oil flow rate: & $3 \leq \dot{m}_{oil} \leq $ \SI{25}{kg/s}         & -               \\
                             & $-1.2 \leq \ddot{m}_{oil} \leq $ \SI{1.2}{kg/s^2}  & -               \\
\bottomrule
\end{tabular}
}
\end{table}


\section{Results and discussion}
\label{sec:results_and_discussion}

This section presents the results of closed-loop simulations of the open
sCO$_2$ cycle detailed in Sec.\;\ref{sec:control_problem} according to the
approach shown in Fig.\;\ref{fig:simulation_setup}.
These simulations use the simulation model, control model, and controller presented in
Secs.\;\ref{sec:simulation_model}, \ref{sec:controller_design}, and
\ref{sec:control_model} respectively.


\subsection{Controller tuning}
\label{sec:controller_tuning}

The controller was manually tuned on the test cases presented in
Sec.\;\ref{sec:closed_loop_simulations}, yielding the weights in
Eq.\;\ref{eq:nominal_q_r} and parameter values in
Tab.\;\ref{tab:controller_settings}.
The controller was tuned according to the following logic.
Plants that can respond quickly to power demand changes are critical for
maintaining power quality and reliability, especially with high supply-side
uncertainty, so net power tracking is weighted highly.
Turbine inlet temperature tracking corresponds to maximizing cycle thermodynamic
efficiency, which brings moderate financial benefit, but does not affect power
system stability, and so thus weighted less highly.
The control input weights were tuned to
(1) give smooth closed-loop response and
(2) prioritize the use of compressor torque for net power tracking, due to it's
predictable and controllable dynamics.
(Due to the typical ranges of each tracked output, the absolute value of the
turbine inlet temperature tracking weight is larger than that for net power.)
\begin{align}
\label{eq:nominal_q_r}
\text{diag}(Q_z) &= [1\text{E}{-3},\; 2\text{E}1]   &&\left( \mbf{z} = [\dot{W}_{net},\, T_{in,\,t}] \right)
\\
\text{diag}(R) &= [2\text{E}2,\; 1\text{E}5]      &&\left( \mbf{u} = [\dot{T}_m,  \ddot{m}_{oil}] \right) \nonumber
\end{align}

The sampling time, prediction horizon, and control model fidelity (set by the
level of discretization in the fluid streams) were set such that
(1) the controller can adequately control the fastest system dynamics,
(2) the prediction horizon covers a large portion of a typical transient, and
(3) the controller is computationally tractable.
For the chosen settings, control updates are reliably computed in less than
\SI{0.03}{s} (10\% of the sampling time) using a 2015 Intel\textregistered\
Core\texttrademark\ i7-4770.

\begin{table}[!h]
\renewcommand{\arraystretch}{0.9}
\linespread{1.0}\selectfont\centering
{\footnotesize \caption{Controller settings\label{tab:controller_settings}}
\begin{tabular}{lrr}
\toprule
\textbf{Parameter}                                                    & \textbf{Value}                \\
\midrule
Sampling time $\Delta t$:                                             & \SI{0.3}{s}                   \\
Estimator sampling time $\Delta t_e$:                                 & \SI{0.06}{s}                  \\
Prediction horizon $H_p$:                                             & 30 steps (\SI{9}{s})          \\
Heat exchanger cells (control model) $N_{cells,\,HX}$:                & 15 cells                      \\
Pipe cells (control model) $N_{cells,\,pipes}$:                       & 5 cells                       \\
MPC weighting matrices $Q$, $P$, $R$:                               & See Eq.\;\ref{eq:nominal_q_r} \\
\bottomrule
\end{tabular}
}
\end{table}


\subsection{Closed-loop simulations}
\label{sec:closed_loop_simulations}

This section evaluates the controller's performance through closed-loop
simulations.
In these simulations, for each target power output $\dot{W}_{net,\,ref}$, the
turbine inlet temperature setpoint $T_{in,\,t,\,ref}$ is set as the maximum
steady-state turbine inlet temperature, as dictated by system constraints (see
Sec.\;\ref{sec:constraints}).

\begin{figure*}
\vspace{-20pt}
\centering
    \begin{subfigure}[t]{0.499\textwidth}
    \centering
    \includegraphics[width=\textwidth]{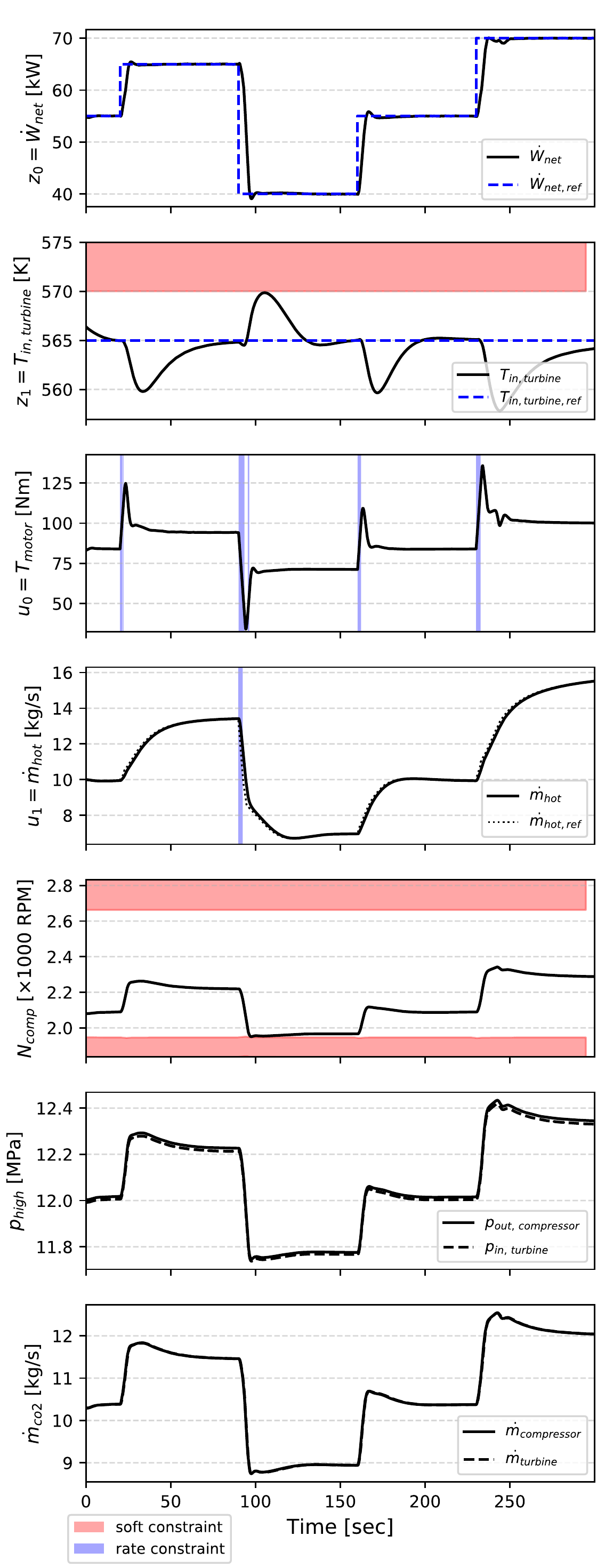}
    \caption{Closed-loop simulation, operation near design-point\label{fig:results_design_point}}
    \end{subfigure}%
    \hfill
    \begin{subfigure}[t]{0.499\textwidth}
    \centering
    \includegraphics[width=\textwidth]{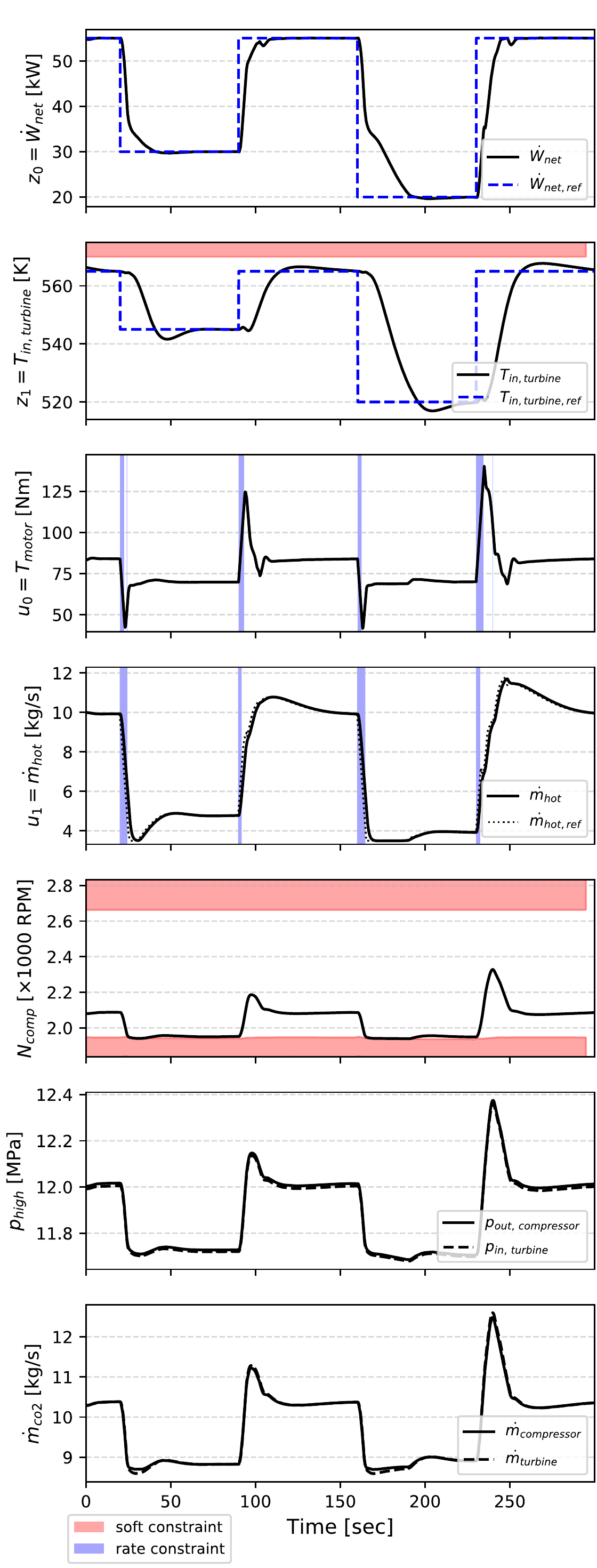}
    \caption{Closed-loop simulation, highly off-design operation\label{fig:results_off_design}}
    \end{subfigure}
\caption{Closed-loop simulations}
\end{figure*}

The first set of simulations (Fig.\;\ref{fig:results_design_point}) show
closed-loop performance near the design-point power output.
In this region, the system can achieve the target net power outputs at the
design-point turbine inlet temperature of \SI{565}{K}.
The core system dynamics are weakly coupled near the design point as the
controller can quickly manipulate compressor torque to enact output power
changes, while more gradually manipulating thermal oil flow rate to track turbine
inlet temperature.

Near the design-point (Fig.\;\ref{fig:results_design_point}), the controller
tracks net power output setpoints quickly, with no steady-state error, and with
negligible overshoots.
The controller enacts fast load changes by aggressively manipulating compressor
torque. 
For example, in response to the load reduction, the controller first decreases
compressor torque at its rate constraint then quickly increases it again to
quickly attain the new target power output without overshooting.
The controller also tracks turbine inlet temperature setpoints well, but less
tightly than for net power, as expected due to it's tuning (see
Sec.\;\ref{sec:controller_tuning}) and the slower dynamics of temperatures
(compared to mass flow rates).
The controller maintains turbine inlet temperature within \SI{10}{K} of the
design value despite operating between approximately 60 to 105\% of nominal
power output.
The compressor surge and turbine inlet temperature constraints are handled well,
with virtually no violations of either.
Closed-loop performance is similar over the entire tested operating range.

The second set of simulations (see Fig.\;\ref{fig:results_off_design}) show
highly off-design operation, with target power outputs far below the nominal
value.
To reach low power outputs (below 60\% of nominal) without violating the surge
constraint, the turbine inlet temperature target must be reduced below its
design value.
Varying the turbine inlet temperature target makes control challenging for the
following reasons:
(1) thermal transients are highly nonlinear and are coupled with CO$_2$ mass
flow (see Sec.\;\ref{sec:mass_flow_rate_and_pressure_modelling});
(2) the compressor operates at the surge boundary for extended periods, limiting
possible control actions; and
(3) the turbine operates over more of its performance map.

For the off-design case (Fig.\;\ref{fig:results_off_design}), the controller
still performs well, despite the wide and nonlinear range of operating points
covered.
Load increases are fast, with average ramp rates during transients comparable to
those achieved near the design point (often in excess of 100\% of nominal output
per minute).
Load reductions take longer as they are limited by the surge constraint.
To significantly reduce net power output, the controller must reduce turbine
inlet temperature to near it's new (lower) target value, involving the
fundamentally slower thermal dynamics of the fluid and component walls.
For all load changes, the controller handles interactions between the thermal
and mass flow transients well, and can deeply converge net power output (the
most important variable operationally) to its target before converging turbine
inlet temperature.
Again, constraint satisfaction is excellent.



The results shown in Figs.\;\ref{fig:results_design_point} and
\ref{fig:results_off_design} support the validity of the approximations used in
the control model, particularly the quasi-steady momentum assumption.
For the open sCO$_2$ cycle, this assumption has two implications:
(1) High-side pressure varies instantaneously with turbine operating point (see
Eq.\;\ref{eq:p_high_update}).
(2) Turbine mass flow rate generally follows compressor mass flow rate according
to a first-order lag model, but deviates during strong thermal transients to
account for the changing specific volume of CO$_2$ (see
Eq.\;\ref{eq:mdot_turb_update}).
The results for both design-point and off-design operation clearly follow these
trends.
Moreover, even during strong thermal transients, compressor and turbine mass
flow rates are very close, suggesting that they may be modeled using a single
variable in the control model.
However, this approach does not provide a mechanism to model mass transfer
between the high- and low-pressure sides of the system, which may be important
when extending the proposed control strategy to closed cycles.


\subsection{Discussion}
\label{sec:discussion}

Control of sCO$_2$ cycles is challenging due to their complex dynamics, which
arise from non-ideal-gas effects and their non-condensing design.
Existing model-based control strategies for similar systems (such as gas
turbines) are not applicable to sCO$_2$ cycles as they do not consider
non-ideal-gas effects.
This article develops a control model for sCO$_2$ cycles by applying
timescale-separation techniques to a high-fidelity truth model,
and by locally linearizing non-ideal-gas turbomachinery performance maps.
This model is theoretically accurate over a wide operating range, subject to
validity of the heat transfer correlations and performance maps.

Overall, the controller performs well for all test cases
supporting the validity of the proposed control strategy.
For MPC, the quality of control depends strongly on the accuracy of the control
model, suggesting that the low-order control model
(Sec.\;\ref{sec:control_model}) approximates the dynamics of the high-fidelity
simulation model (Sec.\;\ref{sec:simulation_model}) reasonably well.
With some modifications --- namely, relaxing the fixed low-side pressure
assumption and adding a valve model --- the control model could be used to
implement MPC for more realistic sCO$_2$ cycle variants, such as the recuperated
cycle or recompression cycle.
The proposed modelling approach may also be applied to similar non-condensing
non-ideal-gas power cycles.

MPC is implemented by linearizing the nonlinear control model online at each
sampling instant.
For this system, obtaining good performance from linear MPC requires that we
retain only certain terms in the linearizations of the wall temperature and
stream internal energy models (see Eqs.\;\ref{eq:wall_temp_update_linear} and
\ref{eq:e_update_linear}).
With this approach, good performance is obtained using linear MPC, even for
large setpoint changes that cover a wide nonlinear operating range.
With linear MPC, the control updates are computed reliably and quickly --- in
less than 10\% of sampling interval on a 2015 Intel\textregistered\
Core\texttrademark\ i7-4770.
It is unclear if linear MPC will still be suitable when considering more complex
cycle configurations with variable low-side pressure and recuperation, however
the our control modelling approach is equally compatible with nonlinear MPC
solvers.

The load changes demonstrated in this article are fast for a thermal power plant.
During design-point and off-design load changes, the controller demonstrates
average ramp rates of approximately 150\% of nominal power output per minute.
There are two caveats to this result: the system is small-scale, and is simpler
than a recuperated or recompression sCO$_2$ cycle, which removes some
challenging dynamics associated with recuperation and variable low-side
pressure.
Real plants may also face additional constraints, such as maximum thermal ramp
rates in component walls or casings.
Slower load changes are expected for more complex cycle configurations and for
real plants, but the encouraging results presented in this article justify
further investigation into the application of MPC to the sCO$_2$ cycle or
related power cycles.

The results presented herein demonstrate several of MPC's principal strengths.
For example, the controller responds rapidly to setpoint changes, coordinating
both control inputs to enact fast and precise changes in net power output.
Despite the system's highly nonlinear dynamics, the controller performs well
over a very wide operating range, covering 35 to 105\% of nominal power output,
without any scheduling of tuning parameters.
The results also highlight MPC's constraint handling capabilities, with the
controller acting on input constraints in response to strong transients and
reliably preventing state constraints from being violated.
The controller can safely track turbine inlet temperature setpoints that lie
close to the maximum turbine inlet temperature, and can reliably operate close
to the compressor surge boundary, thus expanding the safe operating range of the
system.
Many of these qualities directly impact the `flexibility' of the system,
allowing it more quickly respond to load changes and function over a wider and
more efficient operating envelope.
These benefits would likely apply to other non-ideal-gas cycles, suggesting that
MPC might be a useful tool for improving the flexibility of thermal power plants
more generally, thus facilitating higher renewable penetration.








\section{Conclusion}
\label{sec:conclusion}

This article presented a methodology to implement MPC for non-condensing
non-ideal-gas power cycles, then tested this methodology on the high-pressure
side of simple sCO$_2$ cycle power block.
First, to faithfully replicate the system's dynamics, we presented a
high-fidelity gas-dynamics simulation model integrated with empirical
non-ideal-gas turbomachinery submodels.
Next, we developed a control model by applying timescale separation arguments to
the simulation model, then implemented MPC via online relinearization of this
control model.
Finally, we performed closed-loop simulations of a laboratory-scale sCO$_2$ cycle
using the high-fidelity simulation model as a substitute for the real plant.
These simulations demonstrated the effectiveness of the proposed controller for
tracking output power setpoint changes while keeping the system at target
turbine inlet temperatures.
The simulations show that MPC's core strengths, such as good dynamic performance
over a wide operating range and routine handling of constraints, increase the
plant's achievable ramp rates and expand its safe operating envelope.
These flexibility improvements likely apply to other thermal power plants,
suggesting that MPC might be useful for improving power system flexibility more
generally, thus facilitating higher renewable energy share.
The controller is computationally tractable on a standard computer, and with
some modifications, may be extended to more realistic sCO$_2$ cycle
configurations.


\appendix

\section{Performance map models}
\label{sec:performance_maps}

This appendix describes the turbomachinery performance map models used in this
work.
These models are used to compute $\dot{m}_{tb}$ and $T_{out}$ for a turbomachine
given $p_{in}$, $p_{out}$, $T_{in}$, and $N_s$ (as per
Eq.\;\ref{eq:performance_map_reformulated}).
We focus on radial inflow turbines and centrifugal compressors, which are
appropriate for smaller output power sCO$_2$ cycles, such as remote or modular
plants~\citep{jahn_2017}.
Our approach is equally applicable to axial turbomachinery provided that
appropriate maps are used.

\subsection{Turbine}
\label{sec:turbine_map_model}

For typical sCO$_2$ cycle operating conditions, the turbine operates in CO$_2$'s
ideal-gas-like region, so ideal-gas scaling relations can be used to model
off-design turbine operation~\citep{jahn_2017}.
Under this approach~\citep{glassman_1972, whitfield_1990}, turbine performance
is parametrized using the mass flow parameter
\begin{equation}
\label{eq:mfp}
\text{MFP} = \dot{m}_{tb}\, \sqrt{T_{in}} / p_{in}
\end{equation}
and corrected speed
\begin{equation}
\label{eq:n_corr}
N_{cor} = N_s\, \sqrt{T_{in} / T_{in,\,des}},
\end{equation}
where $T_{in,\,des}$ is the design-point inlet temperature.

\begin{figure}[h]
\centering
\includegraphics[width=0.65\columnwidth]{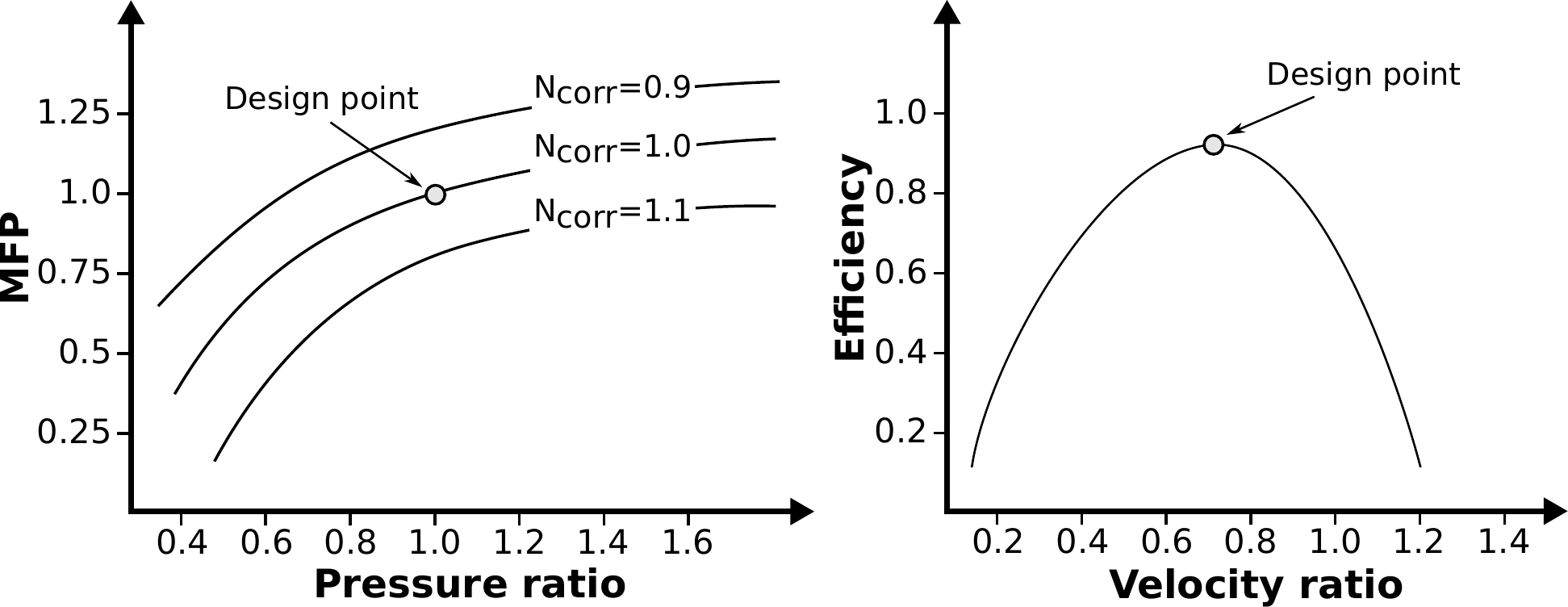}
\caption{Sample turbine performance maps~\citep{glassman_1972, whitfield_1990}}
\label{fig:turbine_performance_maps}
\end{figure}

The design-point performance of a radial inflow turbine operating in the ideal
gas region can be characterized by two maps~\citep{glassman_1972,
whitfield_1990} (shown in Fig.\;\ref{fig:turbine_performance_maps}):
\begin{align}
\label{eq:turb_map_1}
\text{MFP} &= M_{t1}(\text{PR},\, N_{cor})
\\
\label{eq:turb_map_2}
\eta &= M_{t2}(v_{tip}/v_{is}),
\end{align}
where the pressure ratio is PR $= p_{in}/p_{out}$ and
$v_{tip}/v_{is}$ is the ratio of turbine tip speed to the isentropic spouting
velocity of the gas (these two variables lie on the x-axes of the maps).

The turbine performance maps are evaluated as follows.
First, we compute MFP from PR using the map $M_{t1}$ and $N_{cor}$, then we
compute $\dot{m}_{tb}$ from Eq.\;\ref{eq:mfp}.
Next, we compute the isentropic spouting velocity for the current operating
point as
\begin{equation}
\label{eq:v_is}
v_{is} = \sqrt{2\, C_{p,\,in}\, T_{in}\, \left(1 - \text{PR}^{\gamma_{in}/(\gamma_{in}-1)}\right)},
\end{equation}
where the constant pressure and constant volume specific heats are computed from
the inlet state
and the ratio of specific heats is $\gamma_{in} = C_{p,\,in} / C_{v,\,in}$.
We then re-evaluate Eq.\;\ref{eq:v_is} using the design-point pressure ratio
and inlet temperature to get the design-point isentropic spouting velocity
$v_{is,\,des}$.
Next, we compute the velocity ratio as
\begin{equation}
\left.v_{tip}/v_{is}\right. = \left(v_{tip}/v_{is}\right)_{des}\,  \left(N_{cor}/N_{des}\right)\, \left(v_{is}/v_{is,\,des}\right).
\end{equation}
%
From $v_{tip}/v_{is}$, we compute $\eta$ from map $M_{t2}$ then solve for
the outlet state using
\begin{equation}
\label{eq:isentropic_efficiency}
\eta = (h_{in} - h_{out}) / (h_{in} - h_{out,\, is}).
\end{equation}
The remaining outlet properties are then computed using the equation of state.

In this work, the turbine performance maps are based on data
from \citet{hiett_1963}, who published geometries and performance maps for a
range of radial inflow gas turbines developed by Ricardo \& Co.
We use the data for the A70 turbine design due to its geometric similarity to
power generation turbines and similarity in pressure ratio and stage-Mach number
to the turbines employed for sCO$_2$ cycles.


\subsection{Compressor}
\label{sec:compressor_map_model}

\begin{figure}[h!]
\centering
\includegraphics[width=0.65\columnwidth]{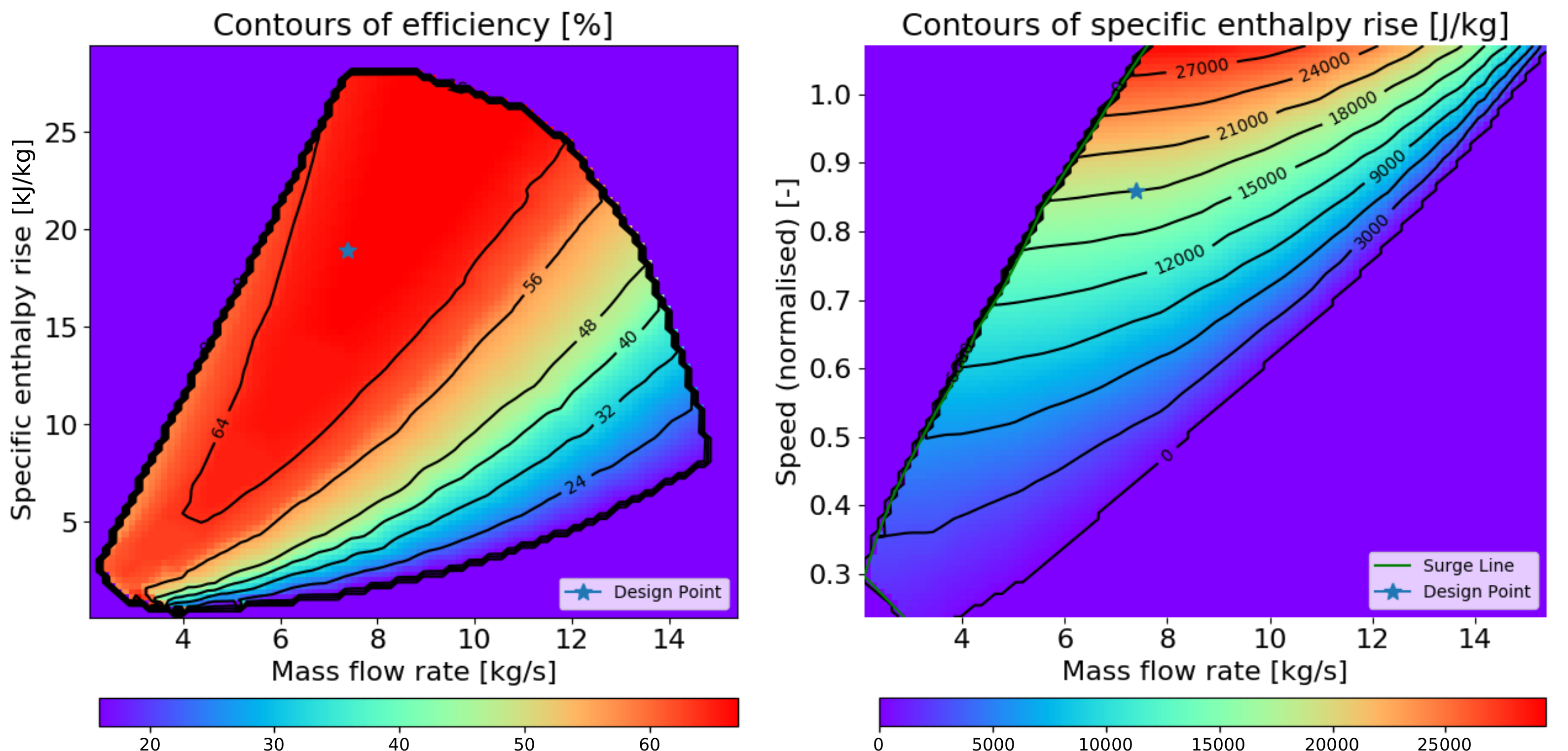}
\caption{
Design-point compressor performance maps~\citep{jahn_2017}.
Off-design performance can be modeled by evaluating the maps with corrected
variables $\dot{m}_{tb,\,cor}$, $N_{cor}$, and $\Delta h_{cor}$.
}
\label{fig:comp_maps}
\end{figure}

In the sCO$_2$ cycle, the compressor inlet conditions are typically close to
CO$_2$'s critical point, where the fluid properties enable efficient
compression.
However, the strong non-ideal-gas property variations in this region cause
compressor performance to change substantially when operating at off-design
inlet conditions.
To capture these non-ideal-gas scaling effects, we use 2D compressor maps that
are parametrized using the corrected variables $\dot{m}_{cor}$,
$N_{cor}$, and $\Delta h_{cor}$~\citep{glassman_1972}.
These corrected variables are computed based on the deviation of the inlet
conditions from some chosen standard conditions ($p_{std}$, $T_{std}$), which
for sCO$_2$ cycles are chosen as CO$_2$'s critical point (\SI{7.366}{MPa} and
\SI{304.1}{K}).
From the standard conditions, the corrected variables are~\citep{glassman_1972}
\begin{align}
\label{eq:mdot_corr}
\dot{m}_{cor} &= \dot{m}_{tb}\, \sqrt{1/V_{cr}}\, \left(p_{std}/p_{in}\right)\, \epsilon
\\
\label{eq:n_s_corr}
N_{cor}            &= N_s\, V_{cr}
\\
\label{eq:h_corr}
\Delta h_{cor}     &= \Delta h\,  V_{cr}
\end{align}
where the non-ideal-gas scaling factors $V_{cr}$ and $\epsilon$ are
\begin{align}
%
&V_{cr} = \left. \left( \gamma_{in,\,std}\, (\gamma_{in}+1)\, T_{std} \right) \middle/ \left( \gamma_{in}\, (\gamma_{in,\,std}+1)\, T_{in} \right) \right.
\\
&\epsilon = \left. \left (\left(\dfrac{2\, \gamma_{in,\,std}}{\gamma_{in,\,std}+1}\right)^{\left(1-\gamma_{in,\,std}^{-1}\right)^{-1}} \right) \middle/
\left( {\dfrac{2\, \gamma_{in}}{\gamma_{in}+1}}^{\left(1-\gamma_{in}^{-1}\right)^{-1}} \right) \right.
\end{align}
Using these corrected variables, compressor performance can be characterized
using two 2D maps (shown in Fig.\;\ref{fig:comp_maps}):
\begin{align}
\label{eq:comp_map_1}
\Delta h_{cor} &= M_{c1}(\dot{m}_{cor},\, N_{cor})
\\
\label{eq:comp_map_2}
\eta &= M_{c2}(\dot{m}_{cor},\, \Delta h_{cor}).
\end{align}

The compressor maps are evaluated in forward mode (where $T_{out}$ and $p_{out}$
are outputs --- see Eq.\;\ref{eq:performance_map}), as follows.
First, $\dot{m}_{tb,\,cor}$ and $N_{cor}$ are computed from the compressor
inlet conditions with Eqs.\;\ref{eq:mdot_corr} and \ref{eq:n_s_corr}.
Then, $\Delta h_{cor}$ and $\eta$ are computed from the maps $M_{c1}$ and
$M_{c2}$.
Finally, $\Delta h$ is computed using Eq.\;\ref{eq:h_corr}, then the outlet
state is computed from $\eta$ using Eq.\;\ref{eq:isentropic_efficiency} (as
done for the turbine).

\textit{Evaluation of compressor model from pressure ratio}



\begin{figure}[h]
\centering
\includegraphics[width=0.50\columnwidth]{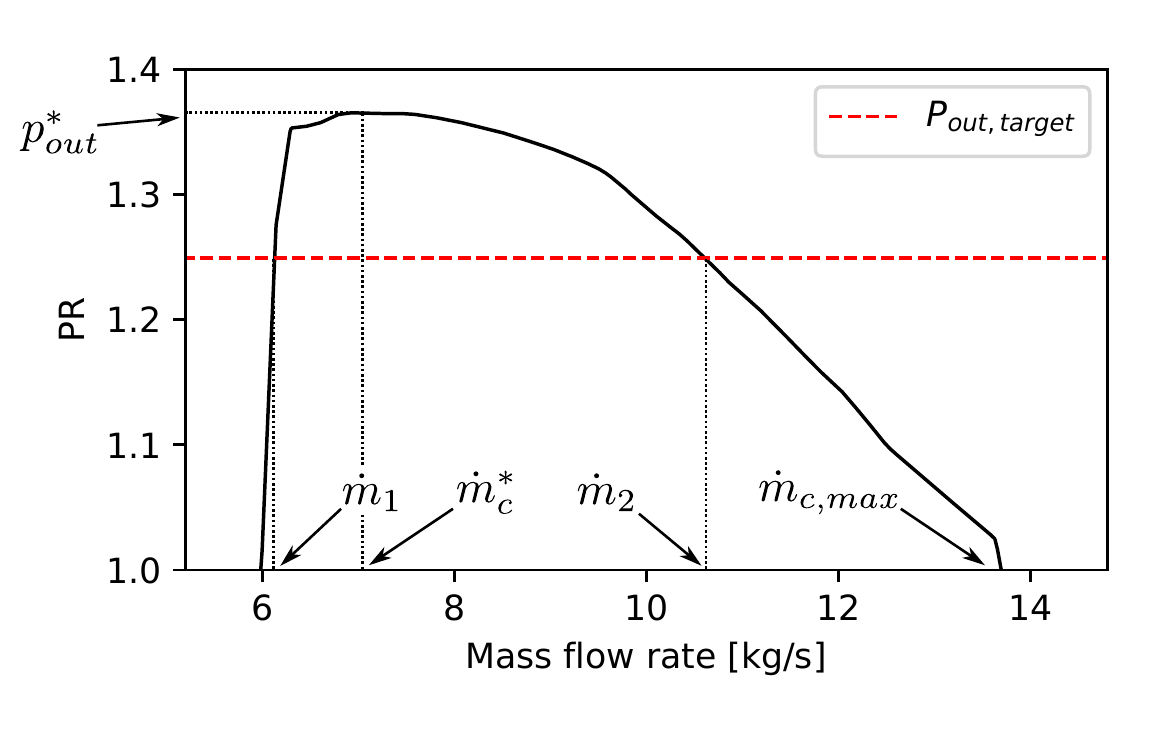}
\caption{
Compressor PR versus mass flow curve for $N_s = 0.916$, $T_{in} = $ \SI{320}{K}, and
$p_{in} = $ \SI{8.629}{MPa}.
The target outlet pressure is \SI{12.5}{MPa}.
}
\label{fig:pr_vs_mdot}
\end{figure}

The compressor performance map model cannot be explicitly reformulated in terms
of pressure ratio.
Accordingly, in simulations, we evaluate this model from inputs $p_{in},\,
p_{out},\, T_{in}$, and $N_s$ (see
Eq.\;\ref{eq:performance_map_reformulated}) by solving the following
root-finding problem (illustrated in Fig.\;\ref{fig:pr_vs_mdot}):
\begin{align}
\label{eq:compressor_root_finding}
\hat{p}_{out},\, \hat{T}_{out} &= f_{tb}(\hat{\dot{m}}_{tb},\, p_{in}, T_{in}, N_s)
\\
0 &= p_{out} - \hat{p}_{out}. \nonumber
\end{align}
In other words, we compute the compressor mass flow $\hat{\dot{m}}_{tb}$ such
that $f_{tb}$ gives the correct $p_{out}$, then take the outlet fluid state as
($p_{out}$, $\hat{T}_{out}$).

As shown in Fig.\;\ref{fig:pr_vs_mdot}, there are two solutions to this
root-finding problem ($\dot{m}_1$ and $\dot{m}_2$).
Due to the surge control that we employ (see
Sec.\;\ref{sec:compressor_speed_constraints}), the compressor is maintained at
the right-side solution $\dot{m}_2$.
We ensure that the root-finder finds this solution by imposing bounds
\begin{equation}
\label{eq:admissible_mdot}
\dot{m}_c^* \leq \hat{\dot{m}}_{tb} \leq \dot{m}_{c,\,max},
\end{equation}
where the surge and maximum mass flow rates are evaluated from lookup tables as
\begin{align}
\label{eq:mc_star}
\dot{m}_c^*       &= f_{c2}(N_{cor}) \\
\dot{m}_{c,\,max} &= f_{c3}(N_{cor}). \nonumber
\end{align}
These lookup tables are formed from data used to build the compressor maps.

We form the compressor maps using data from~\citet{clementoni_2015}.
These maps are linearly scaled by design-point mass flow rate and enthalpy rise
to achieve the power output and pressure ratio.
As primary performance factors (stage Mach number and pressure ratio) can be
maintained through appropriate geometry selections, the scaled maps are
representative of a real machine's behavior.







\bibliography{mpc-for-sco2-cycles}
\bibliographystyle{plainnat}

\end{document}